\documentclass[12pt,a4paper]{article}
\usepackage{epsf,amsmath,amssymb,amsbsy,graphics}
\title{Dynamics of modulated and composite aperiodic crystals: the
       signature of the  inner polarization in the neutron coherent inelastic scattering}
\author{ O. Radulescu$^1$\footnote{{\tt Ovidiu.Radulescu@univ-rennes1.fr }},
  T  Janssen$^2$, and J.Etrillard$^{3,4}$  \\
  $^1$Institut de Recherche Math\'ematique de Rennes, \\
  $^3$Groupe Mati\`ere Condens\'ee et Mat\'eriaux, \\
  University of Rennes 1,  Campus de Beaulieu, \\
          35042 Rennes, France \\
  $^2$ Institute of Theoretical Physics, Nijmegen University,  \\
  Postbus 9010, 6500 GL Nijmegen, The Netherlands \\
  $^4$ Laboratoire L\'eon Brillouin, CEA-CNRS-CE Saclay,
  91191 Gif sur Yvette, France.}

\begin{document}
\maketitle

\begin{abstract} 
We compare within an unifying formalism the dynamical
properties of modulated and composite aperiodic (incommensurate)
crystals.  We discuss the concept of inner polarization and we
define an inner polarization parameter $\beta$ that distinguishes
between different acoustic modes of aperiodic crystals. Although
this concept has its limitations, we show that it can be used to
extract valuable information from neutron coherent inelastic
scattering experiments. Within certain conditions, the ratio
between the dynamic and the static structure factors at various
Bragg peaks depends only on $\beta$. We show how the knowledge of
$\beta$ for modes of an unknown structure can be used to decide
whether the structure is composite or modulated. The same
information can be used to predict scattered intensity within
unexplored regions of the reciprocal space, being thus a guide for
experiments.
\end{abstract}
%
%%%%%%%%%%%%%%%%%%%%%%%%%%%%%%%%%%%%%%%%%%%%%%%%%%%%%%%%%%%%%%%%%%%%%%%%%%
\section{Introduction}\label{sec:intro}
%%%%%%%%%%%%%%%%%%%%%%%%%%%%%%%%%%%%%%%%%%%%%%%%%%%%%%%%%%%%%%%%%%%%%%%%%%
Aperiodic crystals are long-range ordered structures whose
diffraction patterns are made of Bragg peaks. The difference with
respect to normal crystals is that they are not periodic and
one needs $n=3+D$ ($D$ is the
dimension of the internal space) basis vectors to index the
positions of peaks in reciprocal space:

\begin{equation}
{\bf q}=\sum_{i=1}^{n} z_i {\bf q}_i, \,\, z_i \in {\Bbb Z}
\end{equation}

This property has been exploited in crystallography by embedding
the structure in a superspace crystal of dimension $n$. Atomic
positions are the intersections between the D-dimensional atomic
surfaces and the 3-dimensional physical space. For periodic
crystals, the Floquet-Bloch theory reduces the $3N$ dimensional
eigenvector problem of lattice modes ($N$ is the number of atoms)
to a $3N_u$ dimensional problem ($N_u$ being the number of atoms
 inside the unit cell). The similar approach can be used
for aperiodic crystals, but in this case the number of equations
is not reduced; because of incommensurability $N_u=N$.
Nevertheless, as we show in this paper, superspace concepts lead
to important simplifications in the description of modes, that can
be used in the analysis of inelastic neutron scattering
experiments.

At present, three classes of aperiodic crystals are known:
incommensurate modulated crystals, incommensurate composites and
quasicrystals \cite{JanJan80,JanJan87,Cop95,Jan88}. Atomic
surfaces are discontinuous for quasicrystals, but they can be
continuous for modulated crystals and for composites. The borders
between the three classes are not so clear as one may think. The
classification difficulties between modulated and composites
crystals are notorious. If discontinuous atomic surfaces and
occupational modulations are allowed \cite{Pet90,PetMal91}, any
modulated structure can be seen as a composite one and vice-versa.
The choice can be particularly difficult in compounds with a
complex structure. For instance, recent structural neutron
investigations of Bi-2212 superconductors \cite{Etri00} emphasize
the difficulty of deciding whether these materials should be
called composite or modulated. A natural question arises. Can
dynamics as probed by inelastic neutron scattering experiments
\cite{Lovesey,Squires} give an answer to this classification
problem?

In order to answer this question we need to know the typical
dynamical responses of different aperiodic crystals. Several
simplified model systems will be investigated, hoping that more
complex systems behave in resemblance to one of these. Of course,
it is also possible that compounds that structurally are at the
border between the three classes of aperiodic crystals have also
special dynamical properties, at the border between the ones
studied here, in which case the classification problem could be
really intractable.

In this paper, we discuss the linear (small amplitude) excitations
of displacively modulated and composite crystals. The case of
occupational modulations  and that of non-linear (large amplitude)
excitations will be studied elsewhere.

We focus our discussion on the dynamics of low frequency
excitations. The n-dimensional superspace picture suggests that
aperiodic crystals have n branches of propagating hydrodynamic
modes. Only three of those are true Goldstone modes, associated to
the broken translation symmetry of the Hamiltonian. These three
branches are the general acoustic phonons, present in any solid.
The other $p$ branches are called phasons because they are
associated to the broken "phase" symmetry of the ground states
manifold (which is not a symmetry of the Hamiltonian).

The hydrodynamics of composite aperiodic crystals has already been
the object of several theoretical papers
\cite{EmeAxe78,AxeBak82,FinRic83,Cur02}. Although they offer
rather complete results (including damping) these papers suffer
from their generality and from the use of various phenomenological
parameters. It is difficult to extract from them the consequences
of incommensurability on dynamics and to find the answers to
simple (yet important) questions such as which is the contribution
of each mode to the neutron inelastic scattered intensity and
which are the regions in the reciprocal space where these
contributions are most intense. Modes observability is not only
determined by the (integrated) intensity of the peaks, but also by
the line shape. The hotly debated problem of damping, whether
intrinsic as mode-mode coupling \cite{JanRadRub} or extrinsic via
coupling to defects, is left out from our discussion.

The superspace picture suggests that there is a tight connection between
structure and dynamics. From the very beginning of our paper we exploit
this connection and our purpose is to find a small set of quantities
that can be easily interpreted theoretically and measured experimentally,
and that facilitate the classification of aperiodic crystals from the point of view
of their dynamical properties.

%In displacive modulated crystals for each phason branch
%there may also exist a low frequency amplitudon branch,
%corresponding to fluctuations of the modulation amplitude. The
%amplitudon occurs within the soft mode context, when it costs very
%little energy to change the modulation amplitude, i.e. close to
%the incommensurate transition.

\section{Ground states, modulation functions and low frequency excitations}

Throughout this paper we shall deal with 1D models, otherwise we shall stay
as general as possible.

The positions of the atoms, corresponding to a ground state of a
displacively modulated incommensurate crystal, read
\cite{JanJan87}:

\begin{equation}
\begin{split}
& y_{n,g} =  na_1+\delta_g+f_g(na_1+\delta_g) \\
\text{where}\,\, & f_g(y+a_2) = f_g(y)
\end{split}
\end{equation}

The indexes $n \in {\Bbb Z}, g = 1,n_1)$ are for the unit cells of
the basic (non-modulated) structure and for the atoms within this
unit cell, respectively. $f_g$ are periodic modulation functions
of period $a_2$, $a_1$ is the period of the basic structure, the
ratio $\alpha=a_2/a_1$ is irrational.

A composite has several subsystems. We consider here the simplest
case when there are two  subsystems. The positions of atoms
belonging to the two subsystems read \cite{RadJan99}:

\begin{equation}
\begin{split}
& y_{n,g}^{(1)} =  na_1+\delta_g^{(1)}+f_g^{(1)}(na_1+\delta_g^{(1)})  \\
& y_{m,h}^{(2)} =  ma_2+\delta_h^{(2)}+f_h^{(2)}(ma_2+\delta_h^{(2)})  \\
\text{where} \,\, & f_g^{(1)}(y+a_2) = f_g^{(1)}(y) ,\,\,g=1,n_1  \\
\text{and} \,\, & f_h^{(2)}(y+a_1) = f_h^{(2)}(y),\,\,h=1,n_2
\end{split}
\label{eq:groundcomp}
\end{equation}

The modulation function of one subsystem has the periodicity that
the other subsystem had before modulation.

The ground states of incommensurate modulated and composite
crystals are degenerated. This can be explained by the existence
of symmetry groups that leave the configuration energy invariant
\cite{RadJan99}:
\begin{itemize}
\item
Physical space translations, that are uniform displacements.
\begin{eqnarray}
T_{\lambda}(y_{n,g})= & y_{n,g}+ \lambda \label{eq:phonont1} \\
& \notag \\
T_{\lambda}(y_{n,g}^{(1)}, y_{m,h}^{(2)})= & (y_{n,g}^{(1)}+
\lambda, y_{m,h}^{(2)} + \lambda) \label{eq:phonont2}
\end{eqnarray}
where $\lambda \in {\Bbb R} $ \item Discrete inner space
translations, that are combinations of discrete uniform shifts and
relabelling of atomic positions.
\begin{eqnarray}
P_{r,s} (y_{n,g})= & y_{n+r,g} -ra_1  \label{eq:phasont1} \\
&  \notag \\
P_{r,s} (y_{n,g}^{(1)}, y_{m,h}^{(2)})= & (y_{n+r,g}^{(1)} -
ra_1,y_{m+s,h}^{(2)}-ra_1) \label{eq:phasont2}
\end{eqnarray}
where \footnote{the action of $s$ is trivial for modulated
crystals} $r,s \in {\Bbb Z}$.
\end{itemize}

The transformations $P_{r,s}$ change the phase of the modulation
functions and (only for composites)
produce a relative uniform displacement of the subsystems:

\begin{eqnarray}
& P_{r,s} (\{na_1+\delta_g+f_g(na_1+\delta_g)\})= & \notag \\
  & \{ na_1+\delta_g+f_g(na_1+\delta_g+\delta')\} &  \label{eq:pshift1} \\
&& \notag \\
& P_{r,s} (\{na_1+\delta_g^{(1)}+f_g^{(1)}(na_1+\delta_g^{(1)}) , & \notag \\
& na_2+\delta_h^{(2)}+f_h^{(2)}(na_2+\delta_h^{(2)})\})= & \notag \\
 & \{na_1+\delta_g^{(1)}+ f_g^{(1)}(na_1+\delta_g^{(1)}+\delta') ,&  \notag \\
 & na_2+\delta_h^{(2)}-\delta'+f_h^{(2)}(na_2+\delta_h^{(2)}-\delta') \}&  \label{eq:pshift2}
\end{eqnarray}

where the phase variation is:

\begin{equation}
\delta'=ra_1-sa_2 \label{eq:phase}
\end{equation}

\begin{figure}[p ]
\displaywidth\columnwidth \epsfxsize=7truecm
  \centerline{\epsfbox{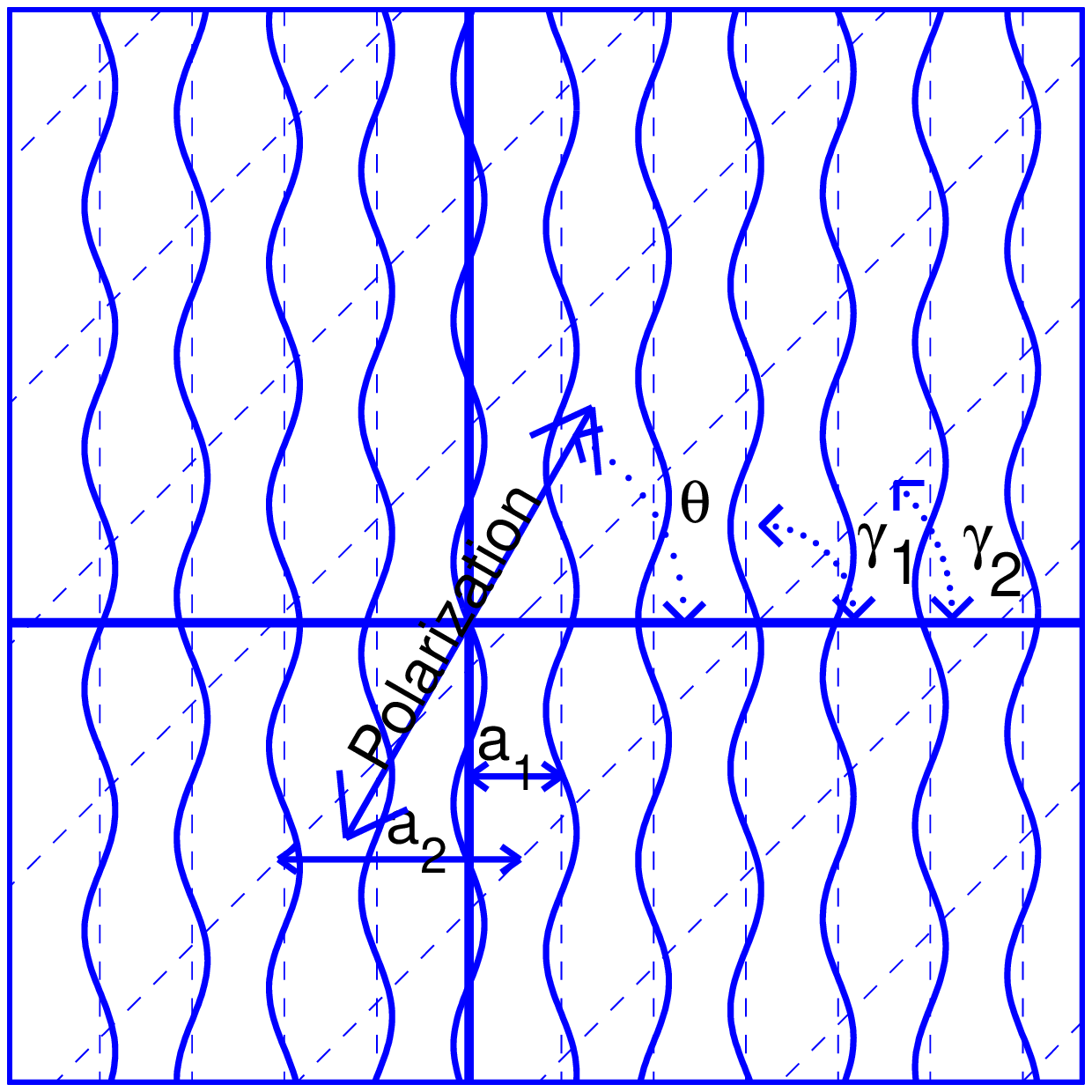} \epsfxsize=7truecm
  \epsfbox{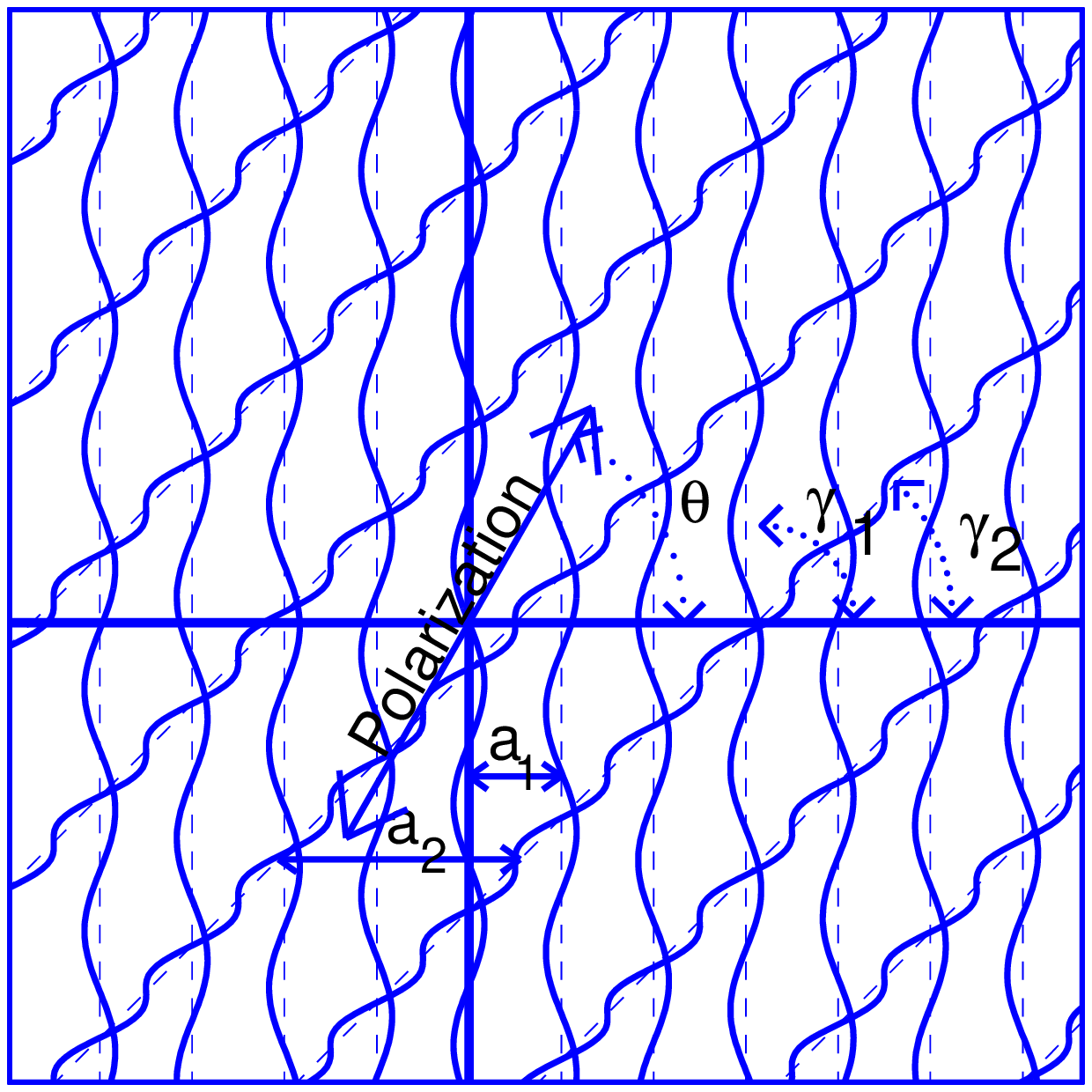 } }

 \centerline{a) \hskip5truecm  b)}
\caption{ Superspace embedding of a) modulated incommensurate
crystals (one set of atomic surfaces) b) composite incommensurate
crystals (two sets of atomic surfaces). For illustration, we have
chosen sinusoidal modulation functions. $\theta$ is the angle made
by the inner polarization and the physical line (horizontal). The
inner polarization parameter is
$\beta=tan(\theta)/[2tan(\gamma_2)-tan(\theta)]$. The acoustical
phonon corresponds to $\theta=0, \beta = 0$. For composites, modes
concentrated on the first subsystem have inner polarization
parallel to the atomic surfaces of the second subsystem, i.e.
$\theta=\gamma_2, \beta = 1$. Modes concentrated on the second
subsystem have inner polarization along the atomic surfaces of the
first subsystem, i.e. $\theta = \pi/2, \beta = -1$.}
\label{fig:superspace}
\end{figure}

Physical space and inner space translations have a simple
geometrical interpretation in a superspace embedding
\cite{RadJan99} where they represent uniform translations of the
superspace crystal. Physical space translations are parallel to
the physical space. Inner space translations make a non-zero angle
$\theta$ with the physical space (Fig.\ref{fig:superspace}). The
physical effect of phase translations depends on $\theta$ but also
on $\gamma_1$,$\gamma_2$, which are the angles made by basic
superspace lattice directions and the physical space. In
Fig.\ref{fig:superspace} these lattice directions define the fixed
repeat distances $a_1$ and $a_2$ along the physical space, while
$\gamma_1$,$\gamma_2$ are free to change. It is convenient to
chose $\gamma_1 = \pi/2$. In this case inner space translations
defined by Eqs. \ref{eq:pshift1},\ref{eq:pshift2} correspond to
$\theta=\pi/2$ (they are orthogonal to the physical space).
%Obviously, the physical meaning of orthogonality of
%translations in superspace depends on the angles
%$\gamma_1$,$\gamma_2$.

More generally, we may define composed translations as
combinations $T_{\lambda} P_{r,s}$ that correspond to translations
in superspace along directions making an angle $\theta$ with the
physical space. If $\gamma_1 = \pi/2$ then one has:

\begin{equation}
\frac{tan(\theta)}{tan(\gamma_2)}=\frac{\delta'}{\lambda}
\label{eq:theta1}
\end{equation}

Because $a_2/a_1$ is irrational, the phase variations $\delta'$
(Eq.\ref{eq:phase}) can approximate with arbitrary precision any
real number (they form a dense set). If the modulation functions
are smooth (the so-called analytic regime \cite{Aub80}) the group
of inner space translations can be extended to a continuous group
$\{P_{\delta'}\}_{\delta'\in {\Bbb R}}$ just by using
Eqs.\ref{eq:pshift1},\ref{eq:pshift2} previously written for
$\delta'=ra_1-sa_2, r,s \in {\Bbb Z}$ and extend them by
continuity to all real values of $\delta'$. Physical space and
inner space translations transform ground states into ground
states. As a consequence of the 2D continuous degeneracy of the
ground states manifold there are two hydrodynamic modes. In the
limit of infinite wavelength $k \rightarrow 0$, the displacements
involved by these modes are the infinitesimal physical space and
inner space translations \cite{RadJan99}:

\begin{itemize}
\item Infinitesimal physical space translations
\begin{eqnarray}
u_{n,g} = \epsilon & \label{eq:phonon1} \\
\{u_{n,g}^{(1)},u_{m,h}^{(2)}\} = \{\epsilon,\epsilon \} &
\label{eq:phonon2}
\end{eqnarray}
\item Infinitesimal inner space translations
\begin{eqnarray}
& u_{n,g} = \eta \frac{df_g}{dx}(na_1+\delta_g) & \label{eq:phason1} \\
& \{u_{n,g}^{(1)},u_{m,h}^{(2)}\} = &   \notag \\
& \{\eta \frac{df_g^{(1)}}{dx}(na_1+\delta_g^{(1)}), -\eta
[1+\frac{df_h^{(2)}}{dx}(ma_2+\delta_h^{(2)})]\} &
\label{eq:phason2}
\end{eqnarray}
\end{itemize}

\begin{figure}[p ]
 \displaywidth\columnwidth \epsfxsize=7truecm
  \centerline{\epsfbox{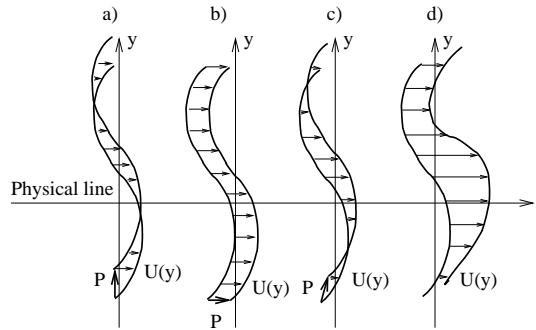} }

\caption{Different inner polarizations $\bf P$ and displacement
hull function $\bf U$ for modulated crystals; a) phason with ${\bf
P} \perp {\bf U}$, b) acoustical phonon ${\bf P} || {\bf U}$ ($\bf
U$ is constant), c) mixed phonon/phason, d) distorted atomic
surface, inner polarization can not be defined. }
\label{fig:polarization}
\end{figure}

Let us define the inner polarization as a vector in superspace:
${\bf P} = (\epsilon,\eta \tan(\gamma_2))$ . A vibration of the
superspace crystal in the direction $\bf P$ will correspond to a
combination of infinitesimal physical space and inner space
translations with coefficients $\epsilon , \eta $
(Fig.\ref{fig:superspace}). The inner polarization should not be
understood as the direction of atom displacements (this is the
physical polarization, always contained in the physical space),
but as a vector belonging to the superspace (abstract construction
describing internal degrees of freedom in a geometrical way). When
the angle $\theta$ between the inner polarization and the physical
space changes, it is only the sequence of signs and magnitudes of
atom displacements that changes, their direction remaining the
same (Fig.\ref{fig:polarization}).

The relation between $\theta$, $\epsilon$, $\eta$ is analogous to
Eq.\ref{eq:theta1}:

\begin{equation}
\frac{tan(\theta)}{tan(\gamma_2)}=\frac{\eta}{\epsilon}
\label{eq:theta2}
\end{equation}

Let us define the polarization parameter $\beta$ as:

\begin{equation}
\beta = \frac{\eta}{2\epsilon - \eta} = \frac{tan (\theta)}
{2\tan(\gamma_2 )- \tan (\theta)} \label{eq:beta}
\end{equation}

In composites $\beta$ may be called "sliding parameter" because at
$k=0,\omega =0$ this is proportional to the relative displacement
between the centers of mass of the subsystems.

\begin{equation}
\beta= \frac{<u_{n,g}^{(1)}>-<u_{m,h}^{(2)}>}
{<u_{n,g}^{(1)}>+<u_{m,h}^{(2)}
>} \label{eq:beta2}
\end{equation}

the averages $<*>$ being with respect to all atoms in a subsystem.

From Eq.\ref{eq:beta2} it follows that in composites the inner
polarization parameter and the participation ratios ${\cal R}_1,{\cal R}_2$ (that
represent the fraction of the total "energy" concentrated on each
subsystem) are related. In the zero$^{\text{th}}$ order of the
modulation amplitude one has:

\begin{equation}
{\cal R}_1=\frac{\alpha <|u^{(1)}_{n,g}|^2>}{\alpha
<|u^{(1)}_{n,g}|^2>+ <|u^{(2)}_{m,h}|^2>} \approx
\frac{\alpha}{\alpha +t^2} \label{eq:particip}
\end{equation}

where $t=\frac{1-\beta}{1+\beta}$, $\alpha=a_2/a_1$.

The modes are concentrated on the first subsystem if ${\cal R}_1=1$,
${\cal R}_2=1-{\cal R}_1=0$. This occurs when $\beta=1$, $\theta=\gamma_2$, thus
when the inner polarization is along the atomic surfaces of the
second subsystem. When $\beta=-1$, $\theta = \pi/2$ (inner
polarization along the atomic surfaces of the first subsystem)
${\cal R}_1=0, {\cal R}_2=1$, hence the modes are concentrated on the second
subsystem.

In the above reasoning we supposed that modulation functions are
smooth and that $\omega=0$. We would like to know if $\bf P$ and
$\beta$ can be defined also elsewhere.

Using  group theory reasonings \cite{Jan79} one may show that
modes in aperiodic crystals are generalized Bloch waves of the type:

\begin{eqnarray}
& u_{n,g,K,y}= & \notag \\
= & \exp[i K (n a_1+\delta_g)]
U_{g,K}[na_1-y/tan(\gamma_2)] & \label{eq:kmodes1} \\
&  \text{for modulated crystals} & \notag
\\
& \{u^{(1)}_{n,g,K,y},u^{(2)}_{m,h,K,y} \}= & \notag \\
= & \{ \exp[i K (n a_1+\delta_g^{(1)})]
U^{(1)}_{g,K}[na_1-y/\tan(\gamma_2)],
& \notag \\
&  \exp[i K (m a_2+\delta_h^{(2)})]
U^{(2)}_{g,K}[ma_2+y/\tan(\gamma_2)]
  \} &  \label{eq:kmodes2} \\
&  \text{for composites} & \notag
\end{eqnarray}

\noindent $K$ is the reduced wave vector
($K=k-2\pi(r/a_1+s/a_2)$), $y$ is the internal space coordinate
(orthogonal to the physical space), $\gamma_2$ is the angle in
Fig.\ref{fig:superspace} (and $\gamma_1=\pi/2$). For each value of
$y$ one has an equivalent realization of the aperiodic crystal, so
one may take $y=0$ in order to obtain the actual displacements.
$U_{g,K},U_{g,K}^{(1)},U_{h,K}^{(2)}$ are periodic functions of
$y$, having periods $a_2,a_2$ and $a_1$, respectively. We shall
call them hull functions and it is useful to picture them in
superspace. At $K=0$, the hull functions play in dynamics the same
r\^ole as the modulation functions play in statics, i.e. their
intersection with the physical space give the displacements of
atoms (Fig.\ref{fig:polarization}). At $K\neq 0$ the hull function
displacements are multiplied by sinusoidal plane waves of
wavelength $2\pi/K$ (Eqs.\ref{eq:kmodes1},\ref{eq:kmodes2}).
Different possible hull functions are represented for modulated
crystals in Fig.\ref{fig:polarization}. In composites, one should
imagine two sets of atomic surfaces and equal polarization vectors
$\bf P$ for the two. As can be understood from
Fig.\ref{fig:polarization}, the inner polarization can be defined
if and only if the atomic surfaces are not distorted. The atomic
surfaces are undistorted in a vibration mode if the motion can be
described as a rigid motion in superspace. In physical space this
comes down to a motion that does not change the local isomorphism
class (the set of local atomic configurations remains the same).
In the analytic regimes this implies that the hull functions are
of the following form:

\begin{eqnarray}
& U_{g,K}(y)= \epsilon(K) + \eta(K) \frac{df_g}{dx}(y+\delta_g) &
\label{eq:hullcondition1}
\\
& U^{(1)}_{g,K}(y)=\epsilon(K) +\eta(K)
\frac{df^{(1)}_g}{dx}[y+\delta_g^{(1)}]  &
\label{eq:hullcondition2}
\\
& U^{(2)}_{h,K}(y)=\epsilon(K) - \eta(K)-\eta(K)
\frac{df^{(2)}_h}{dx}[y+\delta_g^{(2)}] \label{eq:hullcondition3}
\end{eqnarray}

Eqs.
\ref{eq:hullcondition1},\ref{eq:hullcondition2},\ref{eq:hullcondition3}
are exact for $K=0,\omega=0$ in the analytic regime. We conjecture
that they are fulfilled with good accuracy for small $K,\omega$.

In Sec.\ref{subsec:dcm} we shall check the validity of the
undistorted atomic surfaces hypothesis for different models of composite and modulated
crystals. For the analysis of these models we shall also need a set of practical formulas
to compute parameters $\eta(K),\epsilon(K),\beta(K)$.
If Eqs.
\ref{eq:hullcondition1},\ref{eq:hullcondition2},\ref{eq:hullcondition3}
are valid then one has:

\begin{eqnarray}
 \epsilon(K) & = <U_{g,K} (na_1)>  & \label{eq:epsilonk}\\
 \eta(K) & = \frac{< U_{g,K} (na_1) \frac{df_g}{dx}(na_1+\delta_g)>}
 {<(\frac{df_g}{dx})^2>} & \label{eq:etak} \\
 \beta (K) & = \frac{\eta(K) } {2\epsilon(K)-\eta(K)} = & \notag
 \\
  = & \frac{< U_{g,K} (na_1) \frac{df_g}{dx}(na_1+\delta_g)>}
 { 2<(\frac{df_g}{dx})^2>< U_{g,K} (na_1)>-< U_{g,K} (na_1) \frac{df_g}{dx}(na_1+\delta_g)> }
 &  \label{eq:betak} \\
 &  \text{for modulated crystals} & \notag \\
 \epsilon(K) & = <U_{g,K}^{(1)}> & \label{eq:epsilonck}\\
 \eta(K) & = <U_{g,K}^{(1)}> - <U_{h,K}^{(2)}> & \label{eq:etack} \\
 \beta (K) & = \frac{\eta(K)}{2\epsilon(K)-\eta(K)} =
 \frac{<U_{g,K}^{(1)}> - <U_{h,K}^{(2)}>}
 {<U_{g,K}^{(1)}> - <U_{h,K}^{(2)}>} & \label{eq:betack} \\
 &  \text{for composite crystals}  & \notag
\end{eqnarray}

Several things are worth to be noticed:
\begin{itemize}
\item After replacing the displacements by the hull function
displacements, i.e. after eliminating the sinusoidal plane wave of
wavelength $2\pi/K$, $\epsilon(K),\eta(K),\beta(K)$ are calculated
in the same way as they were at $K=0$ (compare
Eqs.\ref{eq:theta2},\ref{eq:beta},\ref{eq:beta2} and
Eqs.\ref{eq:betak},\ref{eq:betack}). \item In modulated crystals,
$\epsilon(K)$ is the displacement of the centre of mass and
$\eta(k)$ is the linear regression coefficient between the hull
function displacements and the derivative of the modulation
function. \item In composites, $\epsilon(K)$ is the displacement
of the centre of mass of one subsystem and $\eta(K)$ is the
relative displacement between the centres of mass of the two
subsystems. \item Contrary to the physical polarization which does
not change inside a branch, the inner polarization expresses
phonon/phason coupling and may depend on $K$.
 \item Generally,
Eqs.
\ref{eq:hullcondition1},\ref{eq:hullcondition2},\ref{eq:hullcondition3}
are only approximate (atomic surfaces are distorted) and their sum
of squares errors \cite{DraSmi81} are given by:

\begin{eqnarray}
& SSE(K) = \sum_{n,g} |U_{g,K}(na_1) - \epsilon(K) - & \notag \\
& - \eta(K)\frac{df_g}{dx}(na_1+\delta_g)|^2 & \label{eq:SSE1}
\\
&  \text{for modulated crystals} & \notag \\
& SSE(K) = \sum_{n,g} |U^{(1)}_{g,K}(na_1)-\epsilon(K) - & \notag
\\
& -\eta(K)\frac{df^{(1)}_g}{dx}(na_1+\delta_g^{(1)})|^2 +& \notag \\
& +\sum_{m,h}|U^{(2)}_{h,K}(ma_2)-\epsilon(K) + \eta(K) + & \notag
\\
& \eta(K)\frac{df^{(2)}_h}{dx}(ma_2+\delta_h^{(2)}) |^2 & \label{eq:SSE2} \\
&  \text{for composite crystals} & \notag
\end{eqnarray}

Eqs.\ref{eq:epsilonk}, \ref{eq:etak}, \ref{eq:epsilonck},
\ref{eq:etack} provide least squares regression coefficients
corresponding to minimum $SSE$ (Eqs.\ref{eq:SSE1},\ref{eq:SSE2}).
The inner polarization parameter follows from
Eqs.\ref{eq:betak},\ref{eq:betack}.
\end{itemize}

Traditionally, modes with $\theta=0$ are called {\em acoustical
phonons}. The name emphasizes the fact that these modes are
generic Goldstone modes, occurring in any solid. Modes with
$\theta \neq 0$ are usually called phasons, because they involve
non-zero phase fluctuations of the modulations. We prefer to call
{\em phason} only the mode whose hull function conserves the
center of mass of the solid. In modulated crystals the phason has
$\epsilon=0,\beta=-1,\theta = \pi/2$ corresponding to a vibration
of the superspace crystal orthogonal to the physical space, hence
along the mean atomic surfaces. The atomic displacements have zero
average and conserve the centre  of mass. In composites the phason
has $\beta=(\rho^{(2)}+\rho^{(1)})/(\rho^{(2)}-\rho^{(1)}),
tan(\theta)=(1+\rho^{(1)}/\rho^{(2)})tan(\gamma_2)$
($\{\rho^{(i)}\}_{i=1,2}$ are the densities of the subsystems)
corresponding to a vibration of the superspace crystal in a
direction between the average directions of the two sets of atomic
surfaces. In the average, the atomic displacements of the
subsystems are antiparallel and conserve the overall centre   of
mass. Of course, there are other possible modes corresponding to
arbitrary angles $\theta$. In composites we shall call {\em
sliding modes} all modes with $\eta \neq 0$ (or equivalently
$\theta\neq 0, \beta \neq 0$). Because $\eta \neq 0$, sliding
modes involve a relative shift between the mass centres of the
subsystems. The phason is a particular sliding mode which
conserves the overall centre of mass, but there is a continuous
set of other possible sliding modes for which the overall centre
of mass is not conserved.

The polarization parameter is zero for acoustical phonons (when
$\eta=0$) and may in principle take any non-zero value for mixed
phonon/phason modes. In the analytic regime, there are two
hydrodynamic branches. The hermiticity of the dynamical matrix
imposes the orthogonality of the modes belonging to the two
branches, which reads:

\begin{eqnarray}
&[1+\beta^1(K)][1+\beta^2(K)] + & \notag \\
& + 4 \beta^1(K)\beta^2(K)<(\frac{df}{dx})^2>= 0 &  \notag \\
& \text{for modulated
crystals} & \label{eq:obeta1} \\
 & \{[1+\beta^1(K)][1+\beta^2(K)]+
<(\frac{df^{(1)}}{dx})^2>
 \}\rho^{(1)} + & \notag \\
& + \{[1-\beta^1(K)][1-\beta^2(K)]+
<(\frac{df^{(2)}}{dx})^2>\}\rho^{(2)} = 0 & \notag \\
& \text{for composites} & \label{eq:obeta2}
\end{eqnarray}

In the non-analytic regime, the modulation functions are
discontinuous and the  phason branch has a gap
($\omega^P(K=0)>0$). The acoustical phonon ($\beta=0$) remains the
only acoustical branch, and obviously no coupling is possible
within the frequency gap (small $K$), because there is no phason
there. Nonetheless, the coupling becomes possible for frequencies
$\omega
> \omega^P(0)$, leading to crossover phenomena
(see Sec.\ref{subsec:dcm}). The derivatives of the modulation
functions are singular. Hull functions may also become singular in
this regime and the equations
\ref{eq:hullcondition1},\ref{eq:hullcondition2},\ref{eq:hullcondition3}
should be understood in the sense of generalized functions.
Although this issue deserves further study, it will not be
addressed in this paper.

%\section{Dynamics in the analytic case}\label{sec:analytic}

\section{Inner  polarization and inelastic neutron scattering}\label{subsec:beta}

Lattice dynamics can be investigated by neutron inelastic scattering.

In coherent neutron inelastic scattering, acoustical modes will be
seen as branches issued from main and satellite Bragg reflections,
whose positions in reciprocal space (for both modulated and
composite crystals) are:

\begin{equation}
k_{r,s}=2\pi[r/a_1 + s/a_2]
\end{equation}

Ignoring damping, the one-phonon inelastic coherent scattering
cross-section obeys \cite{Lovesey,Squires}:

\begin{eqnarray}
& \left(\frac{d^2\sigma}{d\Omega d E}\right)^{inel}_{coh} \sim
\sum_{p} \delta( \omega - \omega^p(K))
  \frac{1}{\omega^{p}(K)} \times & \notag \\
&   \times \left|\sum_n \bar{b}_n \exp{[-W_n(k)]} (u_n^{p}(K).k)
\exp(-i k y_n) \right|^2
\end{eqnarray}

where $K$ (already defined in
Eqs.\ref{eq:kmodes1},\ref{eq:kmodes2}) is the relative distance to
a Bragg peak $k_{r,s}$, i.e. $k=k_{r,s}+K$. The index $p$ is for
different modes of angular frequency $\omega^p(K)$ and
displacements $u_n^{p}(K)$ in a normalized mode. $\bar{b}_n$ is
the average scattering length of the $n^{th}$ atom and
$\exp[-W_n(k)]$ is the Debye-Waller factor.

Let us consider that all scattering lengths and Debye-Waller
factors are equal. Then, ignoring the frequency dependence
$1/\omega^p(K)$ and the slowly varying factor $k \,
\exp{[-W_n(k)]}$, the following quantity, that we call {\em
dynamical scattering factor}, represents the contribution of one
mode to the scattered amplitude along the dispersion branch
$\omega = \omega(K)$:

\begin{equation}
DSF(k)= \lim_{N \rightarrow \infty} \frac{1}{\sqrt{N}}
\sum_{n=1}^{N} u_{n} exp(-ik y_n) \label{eq:dsfdef}
\end{equation}

We have considered normal displacements  $\sum_{n=1}^{N} |u_{n}|^2
=1$. The constant $1/\sqrt{N}$ ensures the convergence of the sum.
A pure acoustical phonon mode at $k=0$ has $u_n= 1/\sqrt{N}$ and
$DSF=1$.

$DSF$ contains information on both structure (via
the atomic positions $y_n$)
and dynamics (via the mode displacements $u_n$).

The {\em static structure factor} is :

\begin{equation}
SF(k) = \lim_{N \rightarrow \infty} \frac{1}{N} \sum_{n=1}^{N} exp(-ik y_n)
\label{eq:sfdef}
\end{equation}

In Appendix 1 we have calculated the SF and the DSF close to a
reflection $k_{r,s}$, supposing that
Eqs.\ref{eq:hullcondition1},\ref{eq:hullcondition2},
\ref{eq:hullcondition3} are accurate and that the modulation
functions are smooth. It follows that for small modulation
amplitudes ($\zeta= k_{r,s} \sup|f| << 1$) and for small
reciprocal space distances to the reflection
($\chi=|K/k_{r,s}|<<1$), the ratio DSF/SF depends in lowest order
of $\zeta,\chi$ only on the inner polarization, incommensurability
ratio $\alpha$ and indices r,s (see also \cite{Rad01}).

For modulated crystals one has:

\begin{eqnarray}
& \left( \frac{|DSF(k_{r,s}+K)|}{|SF(k_{r,s})|} \right)^2 = &
\notag \\
& = \frac{1}{1+[1-t(K)]^2<(\frac{df_g}{dx})^2>}\left[ \frac{\alpha
r + t(K) s}{\alpha r +s} \right]^2 & \label{eq:rap1}
\end{eqnarray}

where

\begin{equation}
t(K)= \frac{1-\beta(K)}{1+\beta(K)} \label{eq:tK}
\end{equation}

The corresponding relationship for  composite crystals is:

\begin{equation}
\left( \frac{|DSF(k_{r,s}+K)|}{|SF(k_{r,s})|} \right)^2 =
\frac{n_1\alpha+n_2}{n_1\alpha+n_2 t^2(K)}\left[ \frac{\alpha r +
t(K) s}{\alpha r  + s} \right]^2
\label{eq:rap2}
\end{equation}

where $n_1$,$n_2$ are the numbers of atoms of each subsystem
within the periods $a_1,a_2$, respectively (maximum values of
indexes $g,h$ in Eq.~\ref{eq:groundcomp}).

One should notice that in Eqs.\ref{eq:rap1}-\ref{eq:rap2},
$\frac{|DSF|}{|SF|}$ depends on the interaction details only via
the inner polarization parameter $\beta$. For modulated crystals
$<(\frac{df_g}{dx})^2>$ also depends on the interaction details,
but this quantity necessary for the normalization of the modes is
the same for all reflections $k_{r,s}$. The $K$ dependence of the
ratio $\frac{|DSF|}{|SF|}$ is given by the $K$ dependence of
$\beta$. This simple result can be used to interpret experimental
data, but we should recall that our DSF differs from the
experimental dynamical structure factor by a factor which depends
slowly on $k$ and that we ignored damping.

\section{Dynamics of composite crystals using the double chain model}\label{subsec:dcm}

The double chain model was introduced \cite{RadJan97,RadJan99} in
order to study the dynamics of composite structures made of two
intermodulated subsystems \footnote{A similar model was introduced
in a different physical context \cite{KawMat98}}. This model
consists of two parallel chains of atoms. The atoms interact via
pair potentials and move only longitudinally along the common
direction of the chains.

The Hamiltonian of  the double chain is:

\begin{equation}
\begin{split}
H(p^{(1)}_n,p^{(2)}_m,y^{(1)}_n,y^{(2)}_m)= & \\
= \sum_n [\frac{(p^{(1)}_n)^2}{2m_1}+
\frac{k^{(1)}}{2}(y^{(1)}_{n}-y^{(1)}_{n-1}-a_1)^{2}]+ & \\
+ \sum_m [\frac{(p^{(2)}_m)^2}{2 m_2}+
\frac{k^{(2)}}{2}(y^{(2)}_{m}-y^{(2)}_{m-1}-a_2)^{2}]+
 & \sum_{n,m} V(\frac{y^{(1)}_{n}-y^{(2)}_{m}}{r})
\end{split}
\label{eq:DCM}
\end{equation}

where $k^{(1)}$,$k^{(2)}$ and $m_{1}$, $m_{2}$ are elastic constants and
masses for the two chains, and $V$ is the
interchain potential of range $r$.

\subsection{Analytic regime}

The analytic regime corresponds to weak interaction between the
subsystems, or equivalently to rigid subsystems (large $k^{(1)}$,
$k^{(2)}$), and is characterized by continuous modulation
functions \cite{RadJan99}.

The dynamical scattering factor for the double chain model with a
Gaussian interchain potential

\begin{figure}[p ]
 \displaywidth\columnwidth \epsfxsize=12truecm
  \centerline{\epsfbox{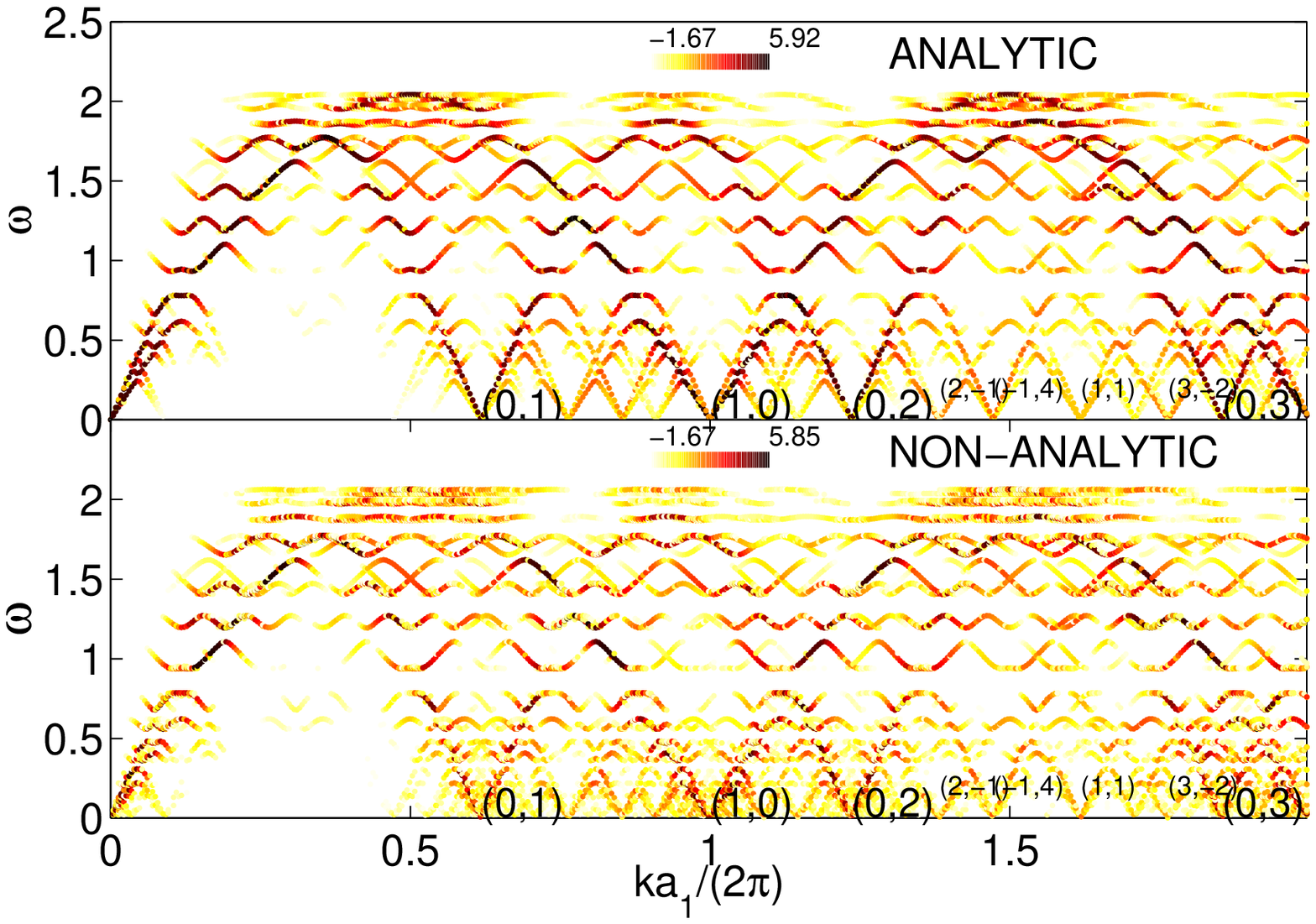 } }

\centerline{a)}

  \displaywidth\columnwidth \epsfxsize=12truecm
  \centerline{\epsfbox{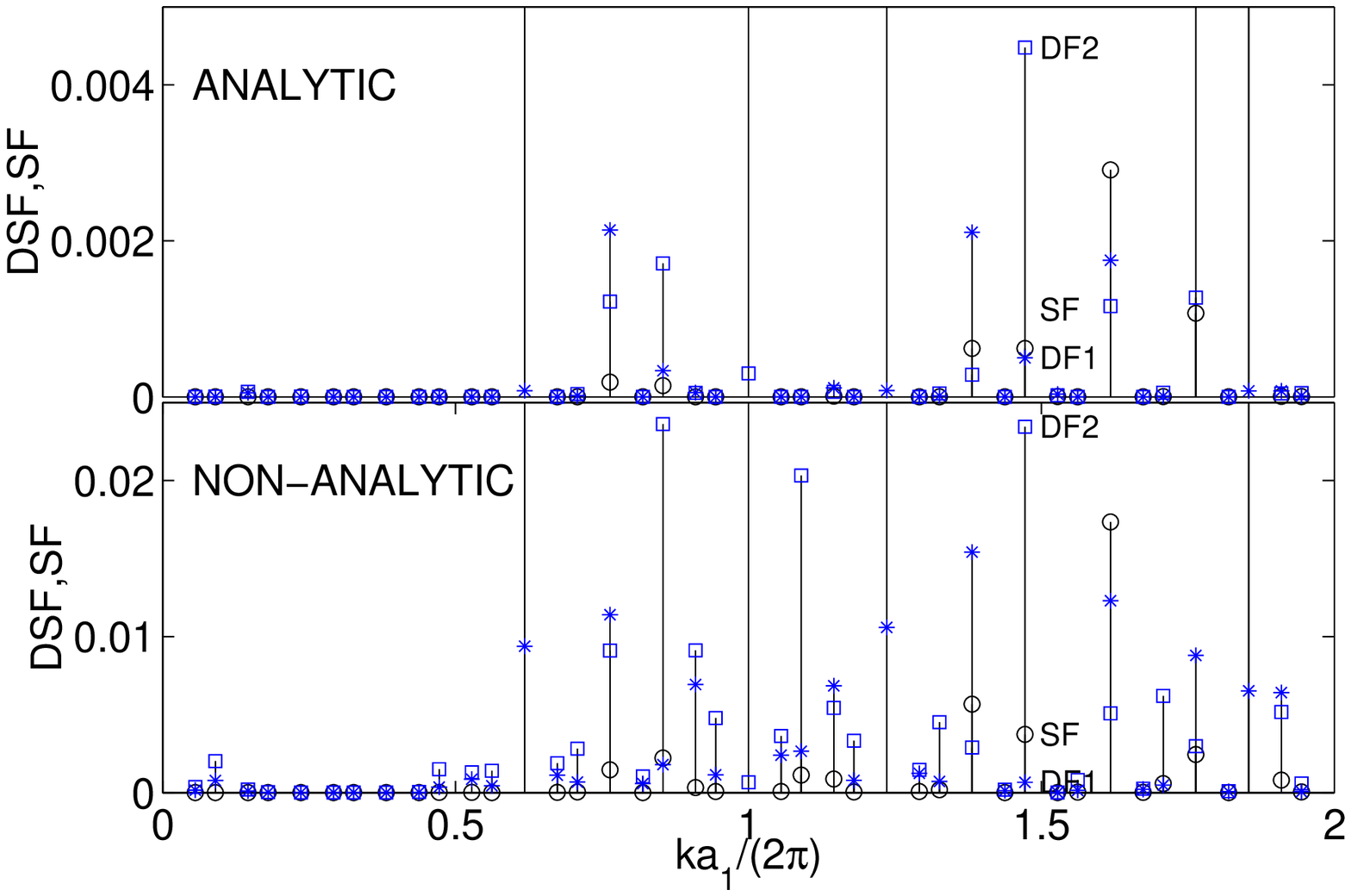 } }

\centerline{b)}

\caption{a) DSF for the double chain model
($\alpha=(1+\sqrt{5})/2$,$m_1=m_2$,$k^{(1)}/k^{(2)}=1.2$, short
range Gaussian potential $r=a_1/3$). The calculations were
performed on approximants of period $a=pa_1=qa_2$, $p=233,q=144$.
Pseudo-colors show DSF values in a logarithmic scale. b) DSF for
the two acoustical modes are compared to SF at different
reflections. $\omega$ is in $\sqrt{k^{(1)}/m_1}$ units}.
\label{fig:DSF}
\end{figure}

($V(x)=-exp(-x^2)$) and equal masses $m_1=m_2$ is shown in
Fig.\ref{fig:DSF}. Several features may be noticed:

\begin{itemize}
\item The dispersion curves  of individual chains, touching the
$\omega=0$ axis at the main Bragg reflections, represent the most
significant features. \item The intensity maxima form an
hierarchical structure. \item From each reflection emerge two
branches. One of the acoustical branches is strong at main Bragg
reflections $(r,0)$ and weak at $(0,s)$, while the other branch
has the opposite behavior. At satellite positions
($(r,s),\,\,r,s,\neq 0$), both branches are weak. \item For
satellites, strongest DSF do not correspond to strongest SF. The
strongest DSF is at $(-1,4)$, while the strongest SF is at
$(1,1)$. \item At high frequency there are nearly flat bands
already observed for the Frenkel-Kontorova model \cite{Luo97} or
for the Fibonacci chain quasicrystal model \cite{Qui96}.
\end{itemize}

%\end{multicols} %%\widetext
%%%%%%%%%%%%%%%%%%%%%%%%%%%%%%%%%%%%%%%%%%%%%%%%%%%%%%%%%%%%%%%%%%%%%%%%%%
%%%%%%%%%%%%%%%%%%%%%%%%%%%%%%%%%%%%%%%%%%%%%%%%%%%%%%%%%%%%%%%%%%%%%%%%%%

In order to understand the low frequency aspect of Fig.
\ref{fig:DSF} one has first to calculate the inner polarization
parameter for the two acoustical branches using Eq.\ref{eq:betak}.

Fig.\ref{fig:Modes} shows the dispersion curves and the inner
polarization parameter $\beta$ for the two acoustic-like branches.
The acoustic-like branch 1 is the pure phonon $\beta = 0,
\theta=0$ at $K=0$ and immediately it becomes a sliding mode
concentrated on the first chain ($\beta=1, \theta=\gamma_2, {\cal
R}_1 = 1$). At $K = 0$,  the acoustic-like branch 2 is the phason
with $\beta = (\rho^{(2)}+\rho^{(1)})/(\rho^{(2)}-\rho^{(1)})= -
\frac{\alpha+1}{\alpha - 1} \approx -4.23$, $\theta=tan^{-1}[tan
(\gamma_2) (1+\rho^{(1)}/\rho^{(2)})]$ and at $K\neq 0$ it becomes
a sliding mode concentrated on the second chain because
$\beta=-1,\theta=\pi/2, {\cal R}_1 = 0 $. The displacements $u_n$
are shown in Fig.\ref{fig:disp}. The jump of $\beta$ very close to
$K=0$ is a consequence of the fact that in our numerical
simulations we use approximants and thus the phason gap is very
small, but not strictly zero.  Strictly speaking the inner
polarization is undefined at $K=0$ because of the degeneracy, but
in practice there is always a small gap that lifts this
degeneracy. Furthermore, although the small K values where the
acoustical phonon lives may be inaccessible to neutron inelastic
scattering, the above remark is important when determining the
speed of sound (this will be $d\omega/dK$ for $K=0,\beta(K)=0$).

The values of the inner polarization parameter explain why near
main Bragg reflections, only one acoustic-like mode is strong (the
one concentrated on the chain producing the Bragg reflection) and
why the other is much weaker (see also Fig.\ref{fig:branch}). The
mode with $\beta=-1$ is visible near reflections $r=0$ and the
mode with $\beta=1$ is visible near reflections $s=0$. This is
expressed by Eq.\ref{eq:rap2} which has been illustrated in
Fig.\ref{fig:ratio}. The above behavior of the inner polarization
depends on the interchain potential and occurs when the slopes of
individual dispersion curves of the two chains are not close one
to another
 ($v_1 = \frac{d\omega_1}{dK} =
 a_1 \sqrt{\frac{k^{(1)}}{m_1}} \neq  v_2= \frac{d \omega_2}{dK}=
 a_2  \sqrt{\frac{k^{(2)}}{m_2}}$).
The same inversion of mode intensities between $(r,0)$ and $(0,s)$
main Bragg reflections can be found in Table.1 of \cite{Bru01} for
the double chain model with a Lennard-Jones potential.

\begin{figure}[p ]
     \displaywidth\columnwidth \epsfxsize=10truecm
   \centerline{\epsfbox{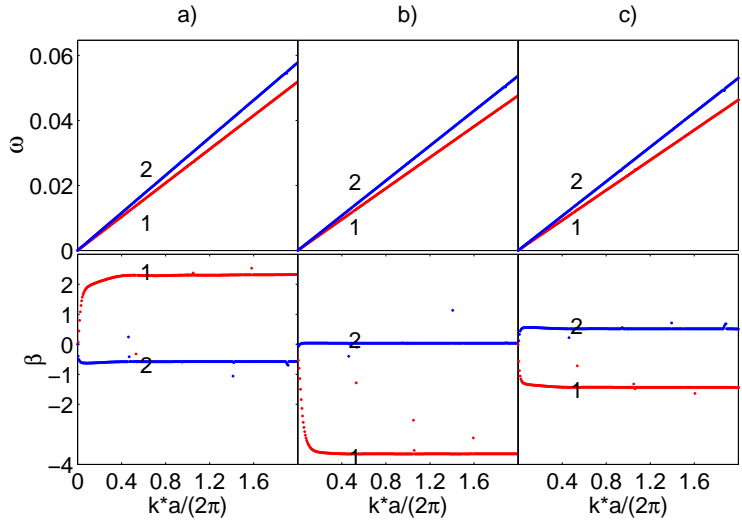}}
 \caption{
 Unfolded dispersion curves and polarization parameters
 in the DCM model, when the single chain sound velocities $v_i=a_i
 \sqrt{k^{(i)}/m_i}, \,i=1,2$ are close one to another.
 a) $v_1=0.9v_2$;
 b) $v_1=v_2$;
 c) $v_1=1.1v_2$;
 }
 \label{fig:equaldcm}
 \end{figure}

When the single chain sound velocities $v_1,v_2$ are close one to
another \footnote{ the meaning of ``how close'' depends on the
type and range of the interchain potential.}, the inner
polarization behaves differently. Single chain modes couple
strongly and tend to be non-concentrated, involving the
participation of both chains ($0 < {\cal R}_1 < 1$ ) and thus
contributing to DSF near both types of main Bragg reflections.
This type of behavior is illustrated in Fig.\ref{fig:equaldcm}
where we have chosen three different situations: $v_1= 0.9 v_2,
v_1=v_2, v_1=1.1 v_2$. Although the single chain sound velocities
are slightly different in the three cases, after the interaction
is set in the dispersion curves look the same; this is the effect
of mode repulsion. Nevertheless, the inner polarization parameters
reach different values in the three situations. In the case
$v_1=v_2$ one of the branches becomes very close to the pure
phason $\beta = -3.58, {\cal R}_1=0.34$ and the other branch is
very close to the acoustical phonon $\beta=0.04, {\cal R}_1 =
0.65$. In the other cases, there are two almost concentrated
sliding modes: $\beta=2.23, {\cal R}_1=0.92$ and $\beta=-0.58,
{\cal R}_1=0.10$ for $v_1= 0.9 v_2$ and $\beta=0.52, {\cal
R}_1=0.94$ and $\beta=-1.42, {\cal R}_1=0.05$ for $v_1= 1.1 v_2$.

\begin{figure}[p ]
     \displaywidth\columnwidth \epsfxsize=10truecm
   \centerline{\epsfbox{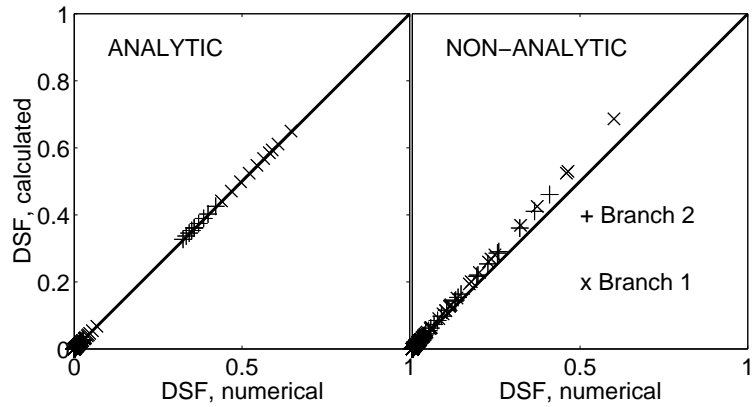} \epsfxsize=10truecm }

 \caption{DCM model: DSF obtained directly from  Eq.\ref{eq:dsfdef} vs. DSF
 calculated from SF via Eq.\ref{eq:rap2} for various main and
 satellite reflections.} \label{fig:compare}
 \end{figure}

\begin{figure}[p ]
   \displaywidth\columnwidth \epsfxsize=10truecm
   \centerline{\epsfbox{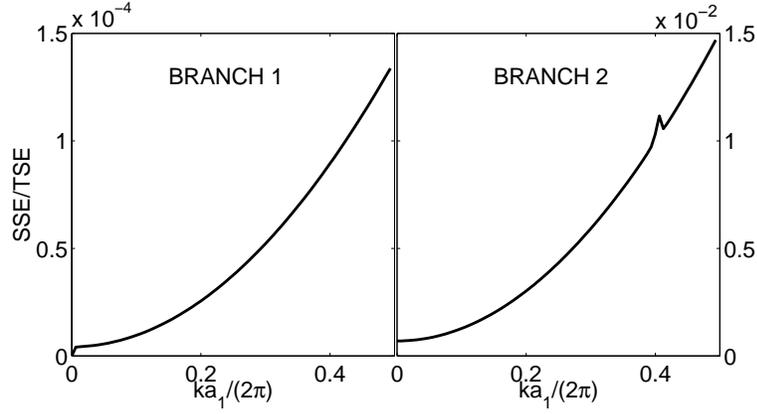} \epsfxsize=10truecm }

 \caption{DCM model: Direct test of
 Eqs.\ref{eq:hullcondition2},\ref{eq:hullcondition3}. The ratio
 SSE/SST for the two acoustical branches in the analytic regime is
 plotted against $K$. The quality of the regression is excellent
 (SSE/SST $<<$ 1 ).} \label{fig:test2}
 \end{figure}

Once SF and $\beta$ known, DSF can be calculated using
Eq.\ref{eq:rap2}. The result is compared to the direct
determination of the DSF (via Eq.\ref{eq:dsfdef}) for wave-vectors
close to two satellite reflections in Fig.\ref{fig:branch}. We
have checked the validity of Eq.\ref{eq:rap2} close to all
reflections with $-10 \leq r,s \leq 10$ and the result is
illustrated in Fig.\ref{fig:compare}. The very good agreement is a
proof that the undistorted atomic surfaces hypothesis is correct
at least for large wavelengths. The errors in Eq.\ref{eq:rap2}
have two origins: eventual deviations from the above hypothesis
and the fact that the calculations leading to Eq.\ref{eq:rap2} are
perturbative and valid for small modulation amplitudes. A direct
test of the undistorted atomic surfaces hypothesis is to check
Eqs.\ref{eq:hullcondition1},
\ref{eq:hullcondition2},\ref{eq:hullcondition3}. This is
illustrated in Fig. \ref{fig:test2} where we have plotted the
ratio SSE/SST against $K$. SSE is the sum of squares error given
by Eq.\ref{eq:SSE2}, while $SST$ is the total sum of squares
estimating the variance of the hull function
($SST=\sum_n[U^{(1)}(na_1)-<U^{(1)}>]^2 +
\sum_m[U^{(2)}(ma_2)-<U^{(2)}>]^2$). $R=\sqrt{1-SSE/SST}$
represents the multiple correlation coefficient ($R$ close to $1$
means good quality of the linear regression \cite{DraSmi81}).

\subsection{Non-analytic regime} \label{subsec:na}

\begin{figure}[p ]
 \displaywidth\columnwidth \epsfxsize=10truecm
  \centerline{\epsfbox{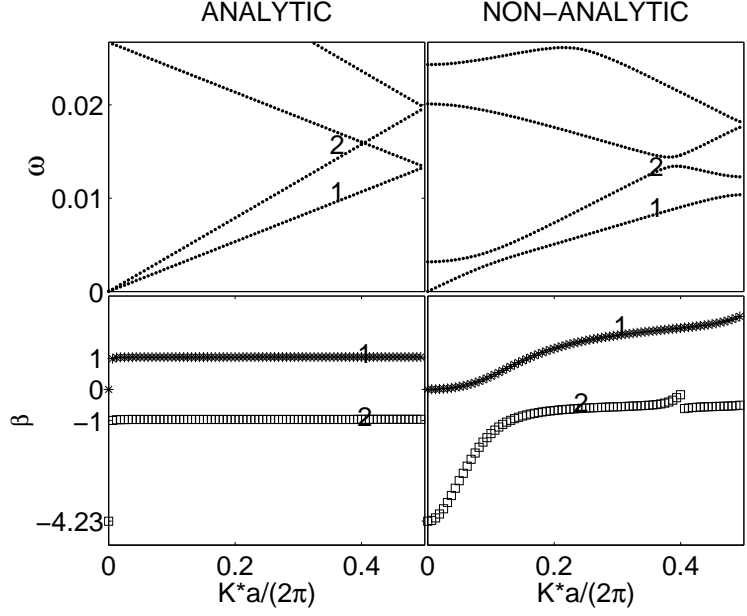}}

\caption{Lowest frequency branches in the double chain model
($\alpha=(1+\sqrt{5})/2$,$m_1=m_2$, $k^{(1)}/k^{(2)}=1.2$, short
range gaussian potential $r=a_1/3$). a) Folded dispersion curves
within the first Brillouin zone of the approximant of period $a=p
a_1 =q a_2$, $p=233,q=144$. $\omega$ is in $\sqrt{k^{(1)}/m_1}$
units; b) Inner polarization parameter $\beta$ along the two
branches. Analytic ($k^{(1)}a^2_1 = 60$): concentrated sliding
modes ($\beta_1=1$,$\beta_2=-1$). Non-analytic ($k^{(1)}a^2_1 =
25.6$): cross-over from non-concentrated modes (the acoustical
phonon $\beta_1 =0$ and the phason $\beta_2 =
-\frac{\alpha+1}{\alpha-1}$) to concentrated sliding modes.  }
\label{fig:Modes}
\end{figure}

\begin{figure}[p ]
\displaywidth\columnwidth \epsfxsize=10truecm
  \centerline{\epsfbox{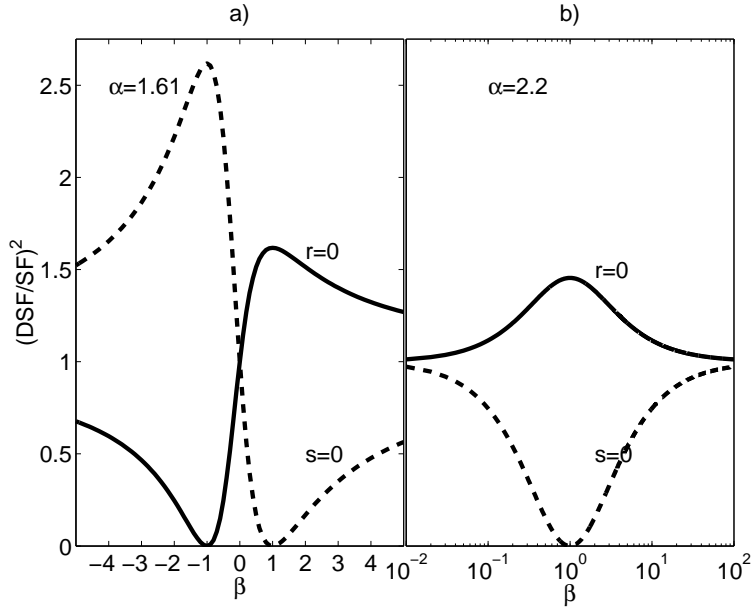} \epsfxsize=15truecm }

\caption{DSF/SF ratio for the  main Bragg reflections of
composites as a function of the inner polarization parameter of the mode.
a) for $\alpha = (1+\sqrt{5})/2$, as in Fig.\ref{fig:Modes};
b) for $\alpha = 2.2$ as in the $Bi-2212$ lamellar superconductor.}
 \label{fig:ratio}
 \end{figure}

\begin{figure}[p ]
    \displaywidth\columnwidth \epsfxsize=10truecm
  \centerline{\epsfbox{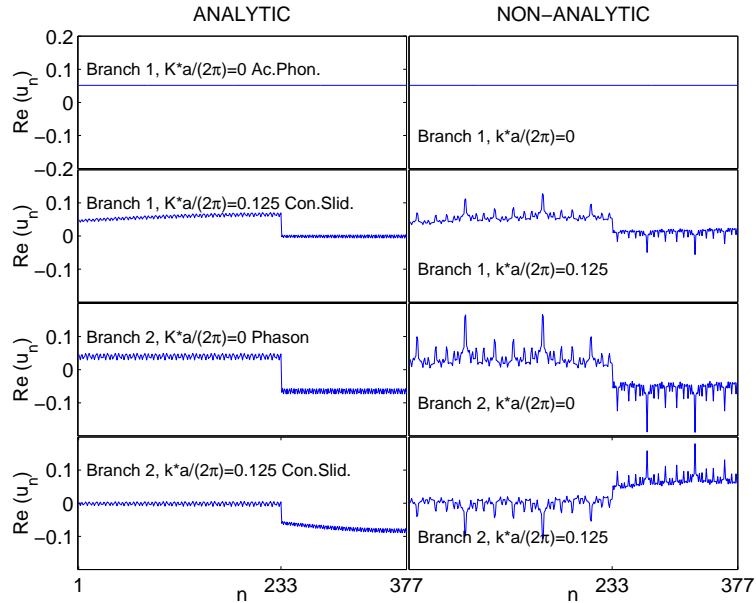} \epsfxsize=15truecm }
\caption{
 Lowest frequency modes in the double chain model.
 Same parameters as for Fig.\ref{fig:Modes}.
 The displacements are longitudinal and correspond to $n=1,233$ for
 the first chain and $n=234,377$ for the second chain. They are uniform for the
 acoustical phonon,
 antiparallel for the phason and
 practically involve only one of the two chains
 for concentrated sliding modes (the other chain presents only
 the small amplitude rapid oscillations equal to the derivative
 of the modulation function). }
 \label{fig:disp}
 \end{figure}

The non-analytic regime occurs when the interaction between the
subsystems is strong, or equivalently when the subsystems are soft
(small $k^{(1)}$, $k^{(2)}$), and is characterized by
discontinuous modulation functions \cite{RadJan99}. As a
consequence, the phason branch starts with non-zero minimum
frequency (gap at $K=0$). This gap can not be noticed in
Fig.\ref{fig:DSF} (it is too small), but it can be seen in
Fig.\ref{fig:Modes}. Several other gaps exist at higher
frequencies and $K \neq 0$.

The variation of $\beta$  inside the two branches is shown in
Fig.\ref{fig:Modes}. For low frequencies the branch 1 has
$\beta=0$, hence it is the pure acoustical phonon, while the
branch 2 is the phason ($\beta =
(\rho^{(2)}+\rho^{(1)})/(\rho^{(2)}-\rho^{(1)})$). This can also
be seen in Fig.\ref{fig:disp} at $K=0$. The branch 1 corresponds
to uniform displacements of the atoms, the branch 2 has the rapid
phason oscillations and a sliding character. For higher
frequencies both branches are sliding modes, almost concentrated
on single chains. Like in the analytic regime, mode concentration
can be avoided when the slopes of the individual dispersion curves
are similar.

The difference with respect to the analytic case is that the
interval of values of $K$ over which the inner polarization
changes is much larger (it scales like the gap). A change of the
slope $d\omega/dK$ accompanies the change of $\beta$
as can be seen in Fig.\ref{fig:Modes}. This could have
consequences when different methods of investigation are employed.
The speed of sound determined by ultrasound and light Brillouin
scattering in the $K=0$ limit measurements may be different from
the slopes $d\omega/dK$ determined by neutron measurements.

Although the static structure factor is different from that in the
analytic regime, the ratio $|DSF/SF|$ is still given with
reasonable accuracy by Eq.\ref{eq:rap2}. This is shown in
Fig.\ref{fig:branch} where the numerical $DSF$, and the $DSF$
calculated from  $SF$ via the Eq.\ref{eq:rap2} are compared for
two satellite reflections. The accuracy of Eq.\ref{eq:rap2} was
tested for all reflections $(r,s)$ with $-10 \leq r,s \leq 10$.
The result is presented in Fig.\ref{fig:compare}.

To summarize, in both analytic and non-analytic regimes, low
frequency features (intensity of acoustical branches close to
different reflections) of $DSF$ depend on the $SF$  and on the
value of the polarization parameter $\beta$. Both $SF$  and
$\beta$ depend on the interaction details of the system,
furthermore $\beta$ may also depend on frequency. For various
model parameters and $K$ values, $\beta$ scans a broad domain of
values, corresponding to inner polarizations of various
inclinations $\theta$ within $(-\pi/2, \pi/2]$.

\section{Dynamics of modulated crystals via the DIFFOUR model}\label{subsec:modulated}

DIFFOUR(discrete frustrated $\Phi^4$) models simulate
\cite{JanTjo} displacively modulated crystals with competing
interactions. A simple example is represented by a chain of atoms
in positions $y_n$ such that the potential energy depends
non-linearly on the fluctuations $x_n=y_n-y_{n-1}-a_1$ of the
first neighbor distances. The Hamiltonian is:

\begin{equation}
\begin{split}
& H(p_n,y_n)=   \\
&= \sum_n \left(\frac{p_n^2}{2m} +  \frac{A}{2} x_n^2 +
\frac{1}{4} x_n^4 - B x_n x_{n-1} + C x_n x_{n-2} \right)
\end{split}
\end{equation}

%One should choose $A< 0$ in order to have double well potentials
%on each site with one positive and one negative minima and $C>0$,
%in order to have competing interactions.

For $A > A^* = B^2/(4C) + 2C$ the ground state is not modulated
($x_n=0$). Modulation occurs for $A \leq A^*$  with a periodicity
$a_2$ that can be incommensurate to the periodicity $a_1$ of the
basic structure: $a_2=2 \pi  a_1 / Q$, where $Q=\arccos
(\frac{B}{4C})$. Close to the commensurate-incommensurate (C-IC)
transition the modulation is analytic ($A \approx A^*$).
Analyticity breaks (the modulation function becomes discontinuous)
when $A$ becomes too small. For $A \approx A^*$, the slopes of the
hydrodynamic branches are given by $\frac{d\omega}{dK}=
\{\frac{1}{m}[(2B+A+C)\cos K - 4 (2C+B) \cos 2K + 9C \cos
3K]\}^{-1/2}$, with $K=0$ for the acoustical phonon and $K=Q$ for
the phason. These slopes are equal $\left( \frac{d\omega}{dK}
\right)_{phonon} = \left( \frac{d\omega}{dK} \right)_{phason} =
3\sqrt{\frac{C}{m}}$ when $B=-2C$.

\begin{figure}[p ]
 \displaywidth\columnwidth \epsfxsize=10truecm
   \centerline{\epsfbox{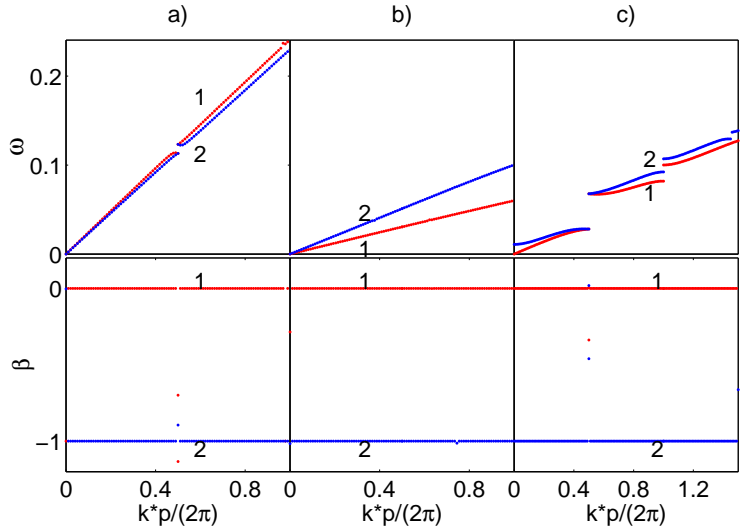}}
 \caption{DIFFOUR model: Unfolded dispersion curves and polarization parameters.
  We used periodic boundary conditions $pa_1=qa_2$.
 a) $C=0.4$, $A=0.99A^*$, $B=-2.1C$, $p=50,q=17$, analytic regime.
 b) $C=0.4$, $A=0.99A^*$, $B=2.5C$, $p=50,q=7$, analytic regime.
 c) $C=0.4$, $A=0.65A^*$, $B=2.5C$, $p=50,q=7$, non-analytic regime.
}
 \label{fig:diffour}
 \end{figure}

\begin{figure}[p ]
    \displaywidth\columnwidth \epsfxsize=10truecm
   \centerline{\epsfbox{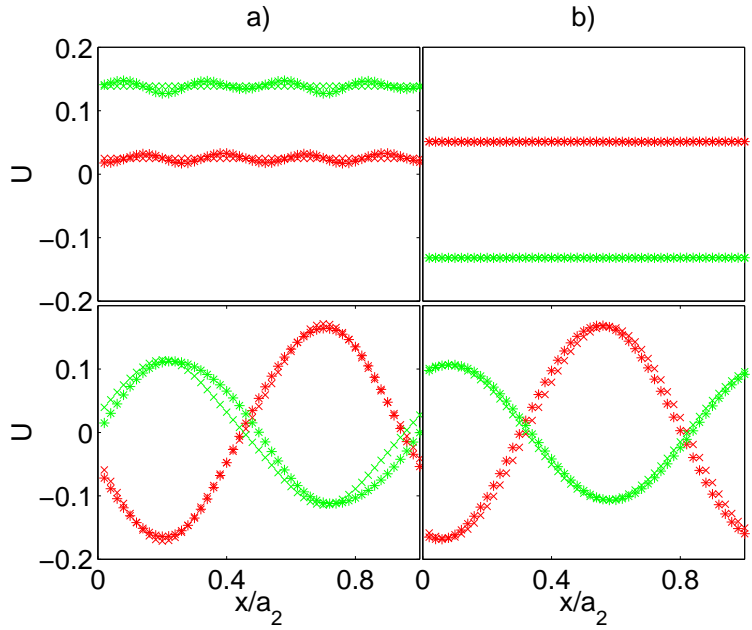}}
 \caption{DIFFOUR model: real and imaginary parts of the
 actual (*) and predicted (x) (by Eq.\ref{eq:hullcondition1}) hull functions
 for the phonon (up) and for the
phason (down) at $K=0.8 \frac{2\pi}{p}$ .
 a) $C=0.4$, $A=0.99A^*$, $B=-2.1C$, $p=50,q=17$ analytic regime.
 b) $C=0.4$, $A=0.99A^*$, $B=2.5C$, $p=50,q=7$ analytic regime.
}
 \label{fig:hull}
 \end{figure}

In Fig.\ref{fig:diffour} we have represented the dispersion curves
and the inner polarization parameter $\beta(K)$ in three
situations: $\left( \frac{d\omega}{dK} \right)_{phonon} \approx
\left( \frac{d\omega}{dK} \right)_{phason}$ for an analytic
modulation, $\left( \frac{d\omega}{dK} \right)_{phonon} \neq
\left( \frac{d\omega}{dK} \right)_{phason}$ for an analytic and a
non-analytic modulation. In all these situations  $\beta$ is
constant and equal to $0$ for the acoustical phonon and is equal
to $-1$ for the phason. No visible phonon/phason coupling occurs.
Further insight is given by the hull functions (Fig.
\ref{fig:hull}). With very good accuracy the phason hull function
is the derivative of the modulation function and has thus
$\beta=-1$. The acoustical phonon corresponds to uniform
displacements of the atoms. A small amplitude modulation (though
with a wavelength that is a fraction of $a_2$, hence different
from the derivative of the modulation function) add to them when
phonon and phason dispersion curves are close
(Fig.~\ref{fig:hull}~a)).

The DIFFOUR model, like the DCM model fit to our general scheme
and satisfies the non-distorted atomic surfaces condition.
Contrary to the DCM model, the values of the inner polarization
parameters are practically restricted to $\beta_1 = 0 $  and
$\beta_2=-1$, corresponding to the acoustical phonon and to the
phason. The following simple argument suggests that close to the
C-IC transition the phason/phonon coupling vanishes  not only for
the DIFFOUR model but for modulated crystals in general.
Considering that the potential energy is $\sum_{m,n}V(y_n-y_m)$,
that close to the transition the modulation is sinusoidal and of
small amplitude, that phonon displacements are $u_n \sim
exp(ikna_1)$ and that phason displacements are $v_n \sim f'(na_1)
exp(ikna_1) \sim exp[i(k+2\pi/a_2)na_1]$, then the phason/phonon
coupling scales like $\sum_{m,n} u^*_n V''(na_1-ma_1) v_m \sim
\sum_m exp(2\pi i m a_1/a_2) = 0 $.

\section{Composites vs. modulated crystals}\label{subsec:kvsm}

Is it possible to distinguish between composites and modulated
crystals by studying their dynamics? A possible experimental
procedure would be to scan DSF in the neighborhood of the main or
satellite reflections. One may choose a set of reflections and
determine the ratio $|DSF/SF|$ for the acoustical modes. From
Eqs.~ \ref{eq:rap1},~\ref{eq:rap2} it follows that the dependence
of the $|DSF/SF|$ ratio on the reflection indexes $r,s$ has the
same form for modulated and composite crystals. The only way to
distinguish between different classes of aperiodic crystals is via
the polarization parameters. We have shown that these are
different for modulated and composite crystals.

First of all, the orthogonality relations
\ref{eq:obeta1},\ref{eq:obeta2} that relate inner polarizations of
the acoustical modes read differently for the two classes. For
instance a $\beta_1 = -1, \theta_1 = \pi/2$ mode (phason in
modulated crystals and concentrated sliding mode in composites) is
accompanied by a $\beta_2=0, \theta_2=0$ (acoustical phonon) in
modulated crystals and by a $\beta_2=1, \theta_2=\gamma_2$ mode
(concentrated sliding mode) in composites. The $\beta_2=0$ and
$\beta_2=1$ modes scatter very differently and this should be seen
in neutron inelastic scattering experiments.

On the other hand, close to the C-IC transition, the inner
polarization of normal modes in modulated crystals is restricted
to $\beta_1=-1$ (phason) and $\beta_2=0$ (acoustical phonon).
Although the numerical simulations using the DIFFOUR model showed
only negligible phason/phonon coupling, the theoretical arguments
that we employed do not exclude the possibility  of a
phason/phonon coupling far from the C-IC transition, for instance
in the non-analytic regime. In composites, various inner
polarizations are possible.

Let us see how the above theoretical considerations can be used to
interpret available experimental results on structure and dynamics
of incommensurate modulated and composite crystals.

Incommensurate modulated crystals have been studied rather
intensively and there are some well characterized model systems
\cite{Cai86,CurJan88,Cum90,CaiEco92}.

For the incommensurate compound $ThBr_4$ \cite{Ber83} the
modulation function is analytic, almost sinusoidal (only first
order satellites are visible). In this case there is no difficulty
to identify the main reflections and the classification task is
trivial. A gapless phason was observed by inelastic neutron
scattering close to the first order satellite $(2, 3, 0.69)$ of
the main Bragg $(2,3,1)$. Despite that the intensity of the
studied satellite reflection is strong, the acoustical phonon
branches were measured only around the main reflections.

In the phase III of biphenyl \cite{Cai86} one can observe strong
satellites, up to order 3 in neutron scattering, suggesting a
non-analytic (soliton) regime or (which is equivalent) a
higher-order commensurate phase. Coherent inelastic neutron
scattering scans around a first-order satellite have revealed a
frequency gap in the phason branch at $K \neq 0$. The presence of
a $K=0$ gap that would be expected in the non-analytic regime has
not been detected, perhaps because of the instrumental resolution
and the heavy damping of these excitations.

There are many structures that can be named "composites". Still, a
precise classification is not established yet.

\begin{figure}[p ]
 \displaywidth\columnwidth
  \centerline{ \epsfxsize=8truecm \epsfbox{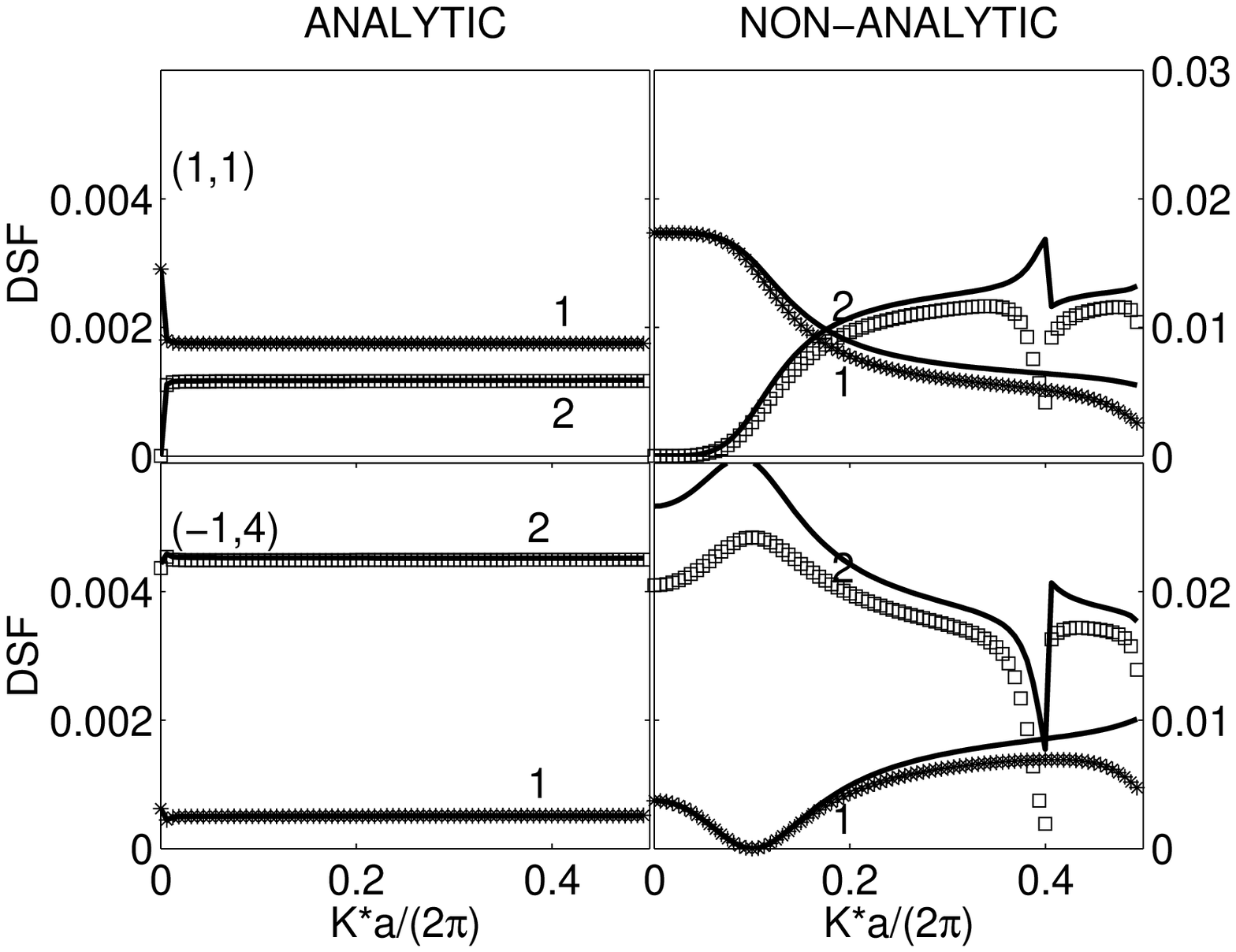}
  \epsfxsize=8truecm \epsfbox{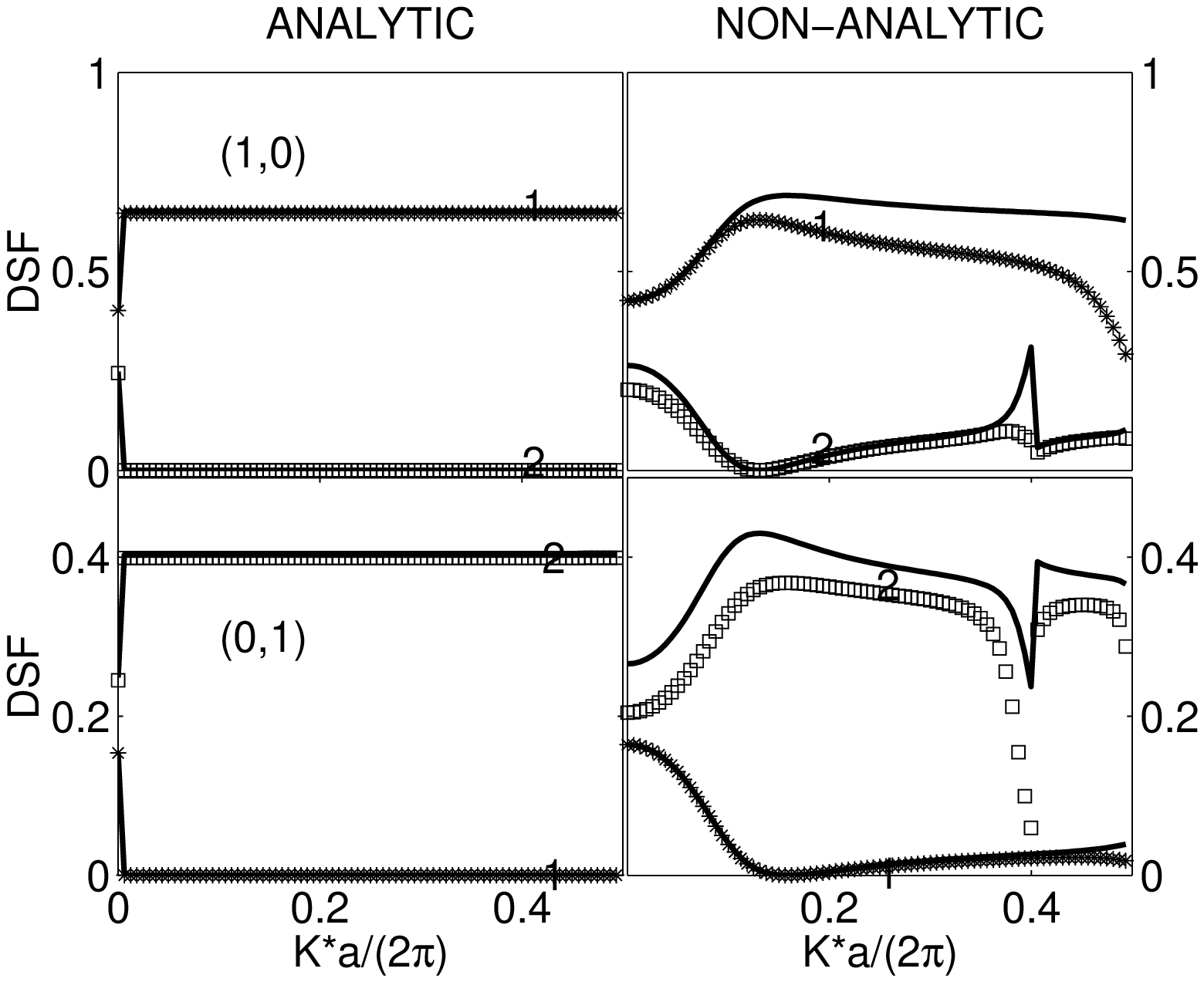}}
\caption{ Numerical values of $DSF$  near the satellites  and main
Bragg reflections, along the low frequency branches, as a function
of the distance in the reciprocal space relative to the
reflection. Thick lines were obtained using  $SF$ and
Eq.\ref{eq:rap2}. Same parameters as for Fig.\ref{fig:Modes}.}
\label{fig:branch}
\end{figure}

The best studied composite system is the mercury chains compound
$Hg_{3-\delta}AsF_6$ with $\delta \approx 0.18$
\cite{Pou78,Hei79}. In this case there is no doubt that the
structure is composite, being made of two weakly interacting
subsystems. The tetrahedra of $AsF_6$ form the host tetragonal
subsystem. There are two mercury subsystems in two orthogonal sets
of channels of the host, running along the equivalent
incommensurate a and b axes of the tetragonal system. At room
temperature the host main Bragg peaks are easily recognizable at
positions $(h,k,l)$ and mercury chains produce diffuse thin sheets
at $(n(3-\delta),k,l)$ orthogonal to $a^*$ and at
$(h,n(3-\delta),l)$ orthogonal to $b^*$. Near main Bragg peaks of
the host and along the incommensurability direction $a$,
acoustical longitudinal modes with a speed
$\frac{d\omega}{dK}=2130 ms^{-1}$ were measured. Near mercury
chain diffuse sheets and along the same direction a gapless
longitudinal branch was measured with a very different speed
$\frac{d\omega}{dK}=3616 ms^{-1}$. According to our discussion
this suggests the existence of two acoustic-like modes with inner
polarization parameters $\beta_1=1$ and $\beta_2=-1$, concentrated
on the host and on the mercury chains, respectively. This picture
is consistent with the weak inter-subsystem coupling and the large
difference between slopes of the dispersion curves
(Figs.\ref{fig:Modes},\ref{fig:branch}).

The alkane/urea inclusion compounds \cite{HolHar96} are also
guest/host type composites with two easily recognizable
subsystems. At all studied temperatures, guest chains (alkane)
have only 1D order and produce diffuse sheets, like in the case of
mercury chains, but the alkane modulation is found to be strongly
non-harmonic \cite{Lef96,Web97}. There are no complete studies of
dynamics for this compound, nevertheless special features of the
longitudinal phonons along the incommensurability direction were
observed near main Bragg reflections of the host by neutron
scattering \cite{Lef98} and confirmed by Brillouin scattering
\cite{Oli97}. These phonons have unusually large damping (approx.
400 GHz, independent of $K$ \cite{Lel00}) compared to the other
acoustical phonons (damping scaling like $K^2$) that could be
explained by a sliding character (finite $\beta$, phason/phonon
coupling). Although this was not studied in this paper, some
authors \cite{FinRic83,Cur02} suggest that damping and sliding are
connected via internal friction. This internal friction can have
an intrinsic origin (occurring for large relative velocities
\cite{JanRadRub}) or an extrinsic one (due to defects).

The case of $Bi-2212$ lamellar superconductor is more complex
because there is no clear separation of the subsystems and the
structural classification is still controversial
\cite{Etri00,Gre01}. Neutron inelastic scattering  \cite{Etri01}
showed the presence of two longitudinal acoustical branches close
to two strong reflections and along the incommensurability
direction. The slopes of the dispersion curves are
$\frac{d\omega}{dK}=2400 \,\, \text{and} \,\, 5900 ms^{-1}$ and
their intensities do not change when one passes from one
reflection to another. According to
Eqs.\ref{eq:rap1},\ref{eq:tK},\ref{eq:rap2}
$(DSF/SF)^2_{(0,s)}/(DSF/SF)^2_{(r,0)} = t^2$ irrespective of the
type of structure, hence $t^2$ should be close to 1 for both
acoustical branches. This is possible only if $|\beta_1|<<1$ and
$|\beta_2|>>1$. One could not speak of a modulated structure,
because this contradicts Eq.\ref{eq:obeta1}. If one of the modes
is the acoustical phonon ($\beta_1 =0, t_1=0$) then in modulated
crystals the other mode must be the phason ($\beta_2 = -1, t_2=
\infty$) being thus invisible close to main Bragg peaks ($s=0$).
In Ref.\cite{Etri00} it was assumed that the structure is
composite and that the two reflections are the main Bragg peaks
$(2,0)$ and $(0,1)$. The intensity ratios are compatible with
Eq.\ref{eq:obeta2} and can be explained by considering that
 one of the longitudinal modes is the
acoustical phonon ($\beta_1 = 0 $) and that the second is the
phason ($|\beta_2| = \frac{\rho^2+\rho^1}{|\rho^2-\rho^1|} =
134.33 >>1$ (this was computed considering that the densities for
the subsystems $Sr_2CaCu_2O_6$ and $Bi_2O_{2.21}$ are $3.248
g/cm^3$ and $3.2 g/cm^3$ respectively). For both the above inner
polarizations the DSF/SF ratio is not changing between the $r=0$
and the $s=0$ Bragg reflections (Fig.\ref{fig:ratio}b)). To
conclude, dynamics suggests that the superconductor is composite,
but the modes are non-concentrated, which is different from the
case of mercury chains compounds. This situation is reproduced by
the DCM model in the rather special situation when $v_1 \sim v_2$
(Fig.\ref{fig:equaldcm}). Before deciding which is really the
case, further study is needed in order to check how the
interaction details (range and type of the inter-chain potential)
and the dimensionality (DCM generalizations to 2D, at least)
influence the inner polarization.
 Referring now to the regime, some of the modulation
functions used in the structure refinement have large amplitudes
and are non-analytic (in particular an occupational modulation is
introduced for the oxygen atoms in the $BiO$ layer of the second
subsystem), while other modulation functions could be analytic.
One may thus think of a situation when the phason gap is below the
experimental resolution of 10 GHz. As discussed in Sec.
\ref{subsec:dcm}, this hypothesis could also explain the observed
discrepancy between the ultrasound measurement of the speed of
sound ($4200 ms^{-1}$) and the slope of the neutron dispersion
curves.

The dynamics of spin ladder compounds
$(Sr_{14-x}Ca_x)Cu_{24}O_{41}$ \cite{Mca88} has been very recently
investigated \cite{BraEtri}.  Again two low frequency longitudinal
modes (along the incommensurability direction)were observed close
to a main Bragg peak of each subsystem $CuO_2$ (chains) and
$(Sr,Ca)_2Cu_2O_3$ (ladders).
 One mode is
acoustic-like (no gap), the other has a gap of $0.35 THz$ (for
$x=0$) and can propagate also transversally. Interestingly enough,
when the latter propagates longitudinally it looses intensity very
rapidly with $K$. Because of the gap, the mode at low frequencies
should be the acoustical phonon. The loss of intensity with $K$ of
the second mode (the phason) suggests that this one tends to
become a sliding mode concentrated on one of the subsystem when
$K$ is increasing, like in Fig.\ref{fig:branch}. The orthogonality
conditions impose that the acoustical phonon should also change
character when $K$ increases, concentrating on the other
subsystem. Our theory predicts that the observation (not yet
performed) of a Bragg peak belonging to the other subsystem could
reveal an opposite behavior of intensity of the acoustical phonon
(this will increase with $K$, see Fig.\ref{fig:branch},
non-analytic regime, (1,0) and (0,1) main Bragg peaks). Of course,
a more cautious analysis should also include damping effects.

%\end{multicols} %\widetext

%\begin{multicols}{2}%%\narrowtext
%%%%%%%%%%%%%%%%%%%%%%%%%%%%%%%%%%%%%%%%%%%%%%%%%%%%%%%%%%%%%%%%%%%%%%

\section{Conclusion and Discussion}

We emphasized the utility of the concept of inner polarization in
the description of low frequency excitations in composite and in
modulated aperiodic crystals. This concept has a meaning for long
wavelengths in the analytic and even weakly non-analytic dynamical
regime, generally for small modulation amplitudes. The inner
polarization is related to internal degrees of freedom of the
aperiodic crystals and should not be looked at as the direction of
atom displacements (this is the physical polarization). Modes
having different inner polarizations differ by the amplitude of
rapid oscillations added to a uniform overall displacement in
modulated crystals and also by the "sliding" character and the
relative participation of the subsystems in composites. The inner
polarization can be defined for modes that do not change the shape
of the superspace atomic surfaces. The possible values of the
inner polarization parameter $\beta$ are different in composites
and modulated crystals.

In modulated crystals, close to the C-IC transition the phason and
the acoustical phonon should be uncoupled.

In composites, the phason and the acoustical phonon can be
coupled. We called ``sliding-modes'' all the modes with $\beta
\neq 0$ (the only exception is the acoustical phonon) because they
produce a relative shift of the mass centres of the subsystems.
The pure phason is a special sliding mode, the only one that
conserves the overall mass centre. One can distinguish two extreme
types of dynamics in composites: concentrated modes (two sliding
modes with $\beta \in \{-1,1\}$, involving each one displacements
of a single subsystem) and non-concentrated modes (the acoustical
phonon, the pure phason or other sliding modes involving both
subsystems). Crossover is possible from one type of dynamics to
another. The variation of the inner polarization and of the
participation ratio with the wave vector $K$ is a different
phenomenon from the crossover between diffusive and propagating
modes discussed in \cite{ZeyFin82,FinRic83,Cur02}.

The inner polarization of the modes influences the neutron
inelastic scattering. For modulated structures the pure phonon is
visible near all main Bragg reflection and satellites, but the
phason is visible only near satellites. For composites, the
``concentrated'' modes are visible only near main Bragg
reflections of the subsystem on which they are concentrated, or
near satellites, ``non-concentrated'' modes are visible near both
types of main Bragg reflections and also near satellites.

The above properties can be used to distinguish between modulated
structures and composites and also between the two extreme types
of dynamics in composites. More information could be (at least in
principle) extracted by an investigation of the satellites. The
inner polarization can be determined from the ratio between the
dynamical structure factor and the statical structure factor at
various reflections. Of course, for more precision one has to
consider also the neglected factor $k \exp[ - W(k)]$ in the $DSF$
and the effect of damping on the line shape.

In experiments,it is important to know for a given mode the set of
satellites where this is visible. A pure phonon scatters strongly
near strong satellites. For modulated structures the empirical
observation is that the same thing is true for the phason. This
can by justified by multiplying eq.\ref{eq:rap1} for
$\beta=-1,t=\infty$ and $k^2 \exp[-2W(k)]$ where $k \sim (r\alpha
+ s)$. We get $I \sim s^2 |SF|^2 exp[-2W(k)]$, where $I$ is the
intensity scattered by the phason. If $|SF|^2$ decreases with $s$
quicker than $1/s^2$, then the first order satellites $s=1$ are
the most intense and correspond to the strongest phason inelastic
scattering. For composites, from eq.\ref{eq:rap2} we get:
\begin{itemize}
\item for $\beta=-1$ (sliding mode concentrated on the second
subsystem) $I \sim s^2 |SF|^2 exp[-2W(k)]$. \item for $\beta=1$
(sliding mode concentrated on the first subsystem) $I \sim r^2
|SF|^2 exp[-2W(k)]$. \item for
$\beta=\frac{\rho^{(2)}+\rho^{(1)}}{\rho^{(2)}-\rho^{(1)}}$ (pure
phason, hence non-concentrated) $I \sim
(r-\frac{\rho^{(1)}}{\alpha \rho^{(2)}}s)^2 |SF|^2 exp[-2W(k)]$.
\end{itemize}
We face here a more complex situation. In composites, one can have
significant values of the $SF$ for high values of $r$ or $s$ for
satellites of the type $(1,s)$ or $(r,1)$ (which are first order
for the second and the first subsystem, respectively). Even if the
$(1,1)$ satellite could be statically more intense, the scattered
intensity may be stronger for $(1,s)$ (for $\beta=-1$) or for
$(r,1)$ (for $\beta=1$). Furthermore, for non-concentrated sliding
modes, there is an extinction rule: the mode will be invisible
close to satellites satisfying $r=\frac{\rho^{(1)}}{\alpha
\rho^{(2)}}s$ (in the DCM model with $m_1=m_2$ this becomes
$r=s$). This type of extinction has been previously reported in
theoretical models \cite{Jan00}. Thus, sliding modes may not give
the strongest inelastic scattering near the strongest satellites.
In a given practical situation one should use Eq.\ref{eq:rap2}
together with the measured values of $|SF|$ in order to calculate
$|DSF|$ and to find out where this is the strongest.

From our discussion of the inner polarization it becomes clear
that sentences of the type "this is the phonon mode and this is
the phason (or sliding) mode" referring to acoustic-like modes in
composites are at least incomplete. Sliding modes may have various
inner polarizations and the pure phonon may simply not exist when
there are two sliding modes.

Further study is needed in order to extend (or to specify the
limits of) the validity of the results presented here, including
the cases of large amplitude and occupational modulations. For a
better analysis of the experimental results a treatment of damping
is necessary.

\section{Appendix }

Let us define the following functions :

\begin{eqnarray}
& \psi_{k,(g,h)}^{(1,2)}(x)=\exp[-ik f^{(1,2)}_{(g,h)}(x)] & \label{eq:function1} \\
& \phi_{k,(g,h)}^{(1,2)}(x)=\frac{df^{(1,2)}_{(g,h)}
}{dx}(x)\exp[-ikf^{(1,2)}_{(g,h)}(x)] & \label{eq:function2}
\end{eqnarray}

$\psi_{k,g}^{(1)}$ and $\phi_{k,g}^{(1)}$ are periodic of period
$a_2$, while $\psi_{k,h}^{(2)}$ and $\phi_{k,h}^{(2)}$ are
periodic of period $a_1$. Let $\tilde{\psi}_{k,(g,h),m}^{(1,2)}$
and $\tilde{\phi}_{k,(g,h),m}^{(1,2)}$ be the Fourier coefficients
of the above functions, such as:

\begin{eqnarray}
& \psi_{k,(g,h)}^{(1,2)} (x) =
\sum_m \tilde{\psi}_{k,(g,h),m}^{(1,2)} \exp(2\pi \,i \, m \, x/a_{(2,1)}) & \\
& \phi_{k,(g,h)}^{(1,2)} (x) =\sum_m
\tilde{\phi}_{k,(g,h),m}^{(1,2)} \exp(2\pi \, i \, m \,
x/a_{(2,1)}) &
\end{eqnarray}

Let us consider periodic approximants such that $a_2/a_1=p/q$,
$p,q$ relatively prime integers. The final result will not depend
on this periodic boundary condition because at the end we shall
impose $p,q \rightarrow \infty$, $a_2/a_1 \rightarrow \alpha$.

In composites, the zero frequency normalized acoustical modes are:

\begin{eqnarray}
& u^{(1)}_{n,g,K}=N_{p,q}^{-1}[1+\beta(K)+2\beta(K)
\frac{df_{g}^{(1)}}{dx}(na_1+\delta_g^{(1)})] \times & \\
& \times \exp [iK(na_1+\delta_g^{(1)})] & \notag \\
& u^{(2)}_{m,h,K}=N_{p,q}^{-1}[1-\beta(K)-2\beta(K)
\frac{df_{h}^{(2)}}{dx}(ma_2+\delta_h^{(2)})] \times & \notag \\
& \times \exp [iK(ma_2+\delta_h^{(2)})] &
\end{eqnarray}

 where

\[
\begin{split}
N^2_{p,q}=n_1
p\{[1+\beta(K)]^2+4\beta^2(K)<|\frac{df_g^{(1)}}{dx}|^2>\}+ &\\
+ n_2 q\{[1-\beta(K)]^2+4\beta^2(K)<|\frac{df_h^{(2)}}{dx}|^2>]\}
&
\end{split}
\] $n_1$,$n_2$ being the number of atoms per each sub-period.
From Eq.\ref{eq:dsfdef} and $k = 2\pi(r/a_1+s/a_2) + K$ one gets:

\begin{equation}
\begin{split}
& DSF^{\text{composites}} (k)= N^{-1}(\alpha,\beta(K)) \{
\sum_{g=1}^{n_1} \alpha [(1+\beta(K))
\tilde\psi_{k,g,s}^{(1)} + \\
& + 2\beta(K)\tilde\Phi^{(1)}_{k,g,s}] +   \sum_{h=1}^{n_2}
[(1-\beta(K)) \tilde\psi_{k,h,r}^{(2)}
-2\beta(K)\tilde\Phi^{(2)}_{k,h,r}] \}
\end{split} \label{eq:dsfap}
\end{equation}

where

\begin{equation}
\begin{split}
 N^2(\alpha,\beta) & =  \lim_{p,q\rightarrow \infty}
\frac{N^2_{p,q}(n_1p+n_2q)}{q^2} = \\ & = (n_1\alpha+ n_2) \{
n_1\alpha [(1+\beta)^2+4\beta^2 <(\frac{df^{(1)}_g}{dx})^2>] +
\\ & + n_2 [(1-\beta)^2+4\beta^2 <(\frac{df^{(2)}_h}{dx})^2>] \}
 \end{split}
 \end{equation}

Similarly, we obtain:

\begin{equation}
SF^{\text{composites}}(k_{r,s}) = \frac{1}{n_1\alpha+n_2} [ \alpha
\sum_{g=1}^{n_1} \tilde{\psi}_{k_{r,s},g,s}^{(1)} +
\sum_{h=1}^{n_2} \tilde{\psi}_{k_{r,s},h,r}^{(2)}] \label{eq:sfap}
\end{equation}

In order to obtain the above results we have used the following
version of Poisson's sum rule:

\begin{equation}
\lim_{p\rightarrow \infty} \frac{1}{p} \sum_{n=1}^{p} f(n a q /p
) \exp(-2\pi \,i n s q/p ) = \tilde{f}_s \label{eq:partial}
\end{equation}

that is valid for any continuous function $f$, periodic of period
$a$. In order to prove Eq.\ref{eq:partial} one could use $f(x)
=\sum_s \tilde{f}_s exp(2 \pi \, i \, s x / a) $ and
$\sum_{n=1}^{p} exp [2 \pi i n m q/p] = p \sum_{k} \delta_{m,kp}$,
valid for $m,p,q$ integers $p,q$ relatively prime.

For small modulation amplitudes $|k f^{(1,2)}|<<1$, we can use the
following linear approximations of
Eqs.\ref{eq:function1},\ref{eq:function2}:

$\psi_{k,(g,h)}^{(1,2)}(x) \approx 1 - i k f^{(1,2)}_{(g,h)}(x)$,
$\phi_{k,(g,h)}^{(1,2)}(x) \approx \frac{df^{(1,2)}_{(g,h)}}{dx}$,
$\tilde{\psi}_{k,g,s}^{(1)} \approx \delta_{s,0} - ik
\tilde{f}_{g,s}^{(1)}$, $\tilde{\phi}_{k,g,s}^{(1)} \approx
\frac{2\pi i s}{a_2} \tilde{f}_{g,s}^{(1)}$,
$\tilde{\psi}_{k,h,r}^{(2)} \approx \delta_{r,0} - ik
\tilde{f}_{h,r}^{(2)}$, $\tilde{\phi}_{k,h,r}^{(2)} \approx
\frac{2\pi i r}{a_1} \tilde{f}_{h,r}^{(2)}$. Substituting in
Eqs.\ref{eq:dsfap},\ref{eq:sfap} one gets :

\begin{eqnarray}
& DSF^{\text{composites}} (k)= N^{-1}(\alpha,\beta(K)) \{ \alpha
n_1 [1+\beta(K)] \delta_{s,0}+ & \notag \\
&+ n_2 [1-\beta(K)]\delta_{r,0} -iK\beta(K)F^-_{r,s,\alpha}-
\notag
\\ & - i
[2\pi(1+\beta(K))\frac{r}{a_1}+2\pi(1-\beta(K))\frac{s}{a_2}+K]F^+_{r,s,\alpha}
\} \label{eq:dsfaplin1} \\
&SF^{\text{composites}}(k_{r,s}) = \frac{1}{n_1\alpha+n_2} [
\alpha n_1 \delta_{s,0}+ n_2 \delta_{r,0} - & \notag \\
& - 2\pi i (\frac{r}{a_1}+\frac{s}{a_2}) F^+_{r,s,\alpha}
\label{eq:sfaplin1}
\end{eqnarray}

where

\begin{equation}
F^\pm_{r,s,\alpha} = \alpha \sum_{g=1}^{n_1} \tilde{f}_{g,s}^{(1)}
\pm \sum_{h=1}^{n_2} \tilde{f}_{h,r}^{(2)} \label{eq:fpm}
\end{equation}

The same kind of arguments were used for modulated crystals to
obtain:

\begin{eqnarray}
& DSF^{\text{modulated}} (k)=
[|\epsilon(K)|^2+|\eta(K)|^2<(\frac{df_g}{dx})^2> ]^{-1/2} \times
& \notag \\
& \times \{ \epsilon(K) \delta_{s,0}
-i[\epsilon(K)(K+2\pi\frac{r}{a_1}) + \notag \\ &
(\epsilon(K)-\eta(K))2\pi\frac{s}{a_2}] <\tilde{f}_{g,s}>
\} \label{eq:dsfaplin2} \\
&SF^{\text{modulated}}(k_{r,s}) =   \delta_{s,0} - 2\pi i
(\frac{r}{a_1}+\frac{s}{a_2})  <\tilde{f}_{g,s}>
\label{eq:sfaplin2}
\end{eqnarray}

where $<\tilde{f}_{g,s}>=\frac{1}{n_1}\sum_{g=1}^{n_1}
\tilde{f}_{g,s}$.

%%%%%%%%%%%%%%%%%%%%%%%%%%%%%%%%%%%%%%%%%%%%%%%%%%%%%%%%%%%%%%%%%%%%%%%
%%%%%%%%%%%%%%%%%%%%%%%%%%%%%%%%%%%%%%%%%%%%%%%%%%%%%%%%%%%%%%%%%%%%%
%%%%%%%%%%%%%%%%%%%%%%%%%%%%%%%%%%%%%%%%%%%%%%%%%%%%%%%%%%%%%%%%%%%%%%

%%%%%%%%%%%%%%%%%%%%%%%%%%%%%%%%%%%%%%%%%%%%%%%%%%%%%%%%%%%%%%%%%%%%%%
%\begin{multicols}{2} %\narrowtext
%%%%%%%%%%%%%%%%%%%%%%%%%%%%%%%%%%%%%%%%%%%%%%%%%%%%%%%%%%%%%%%%%%%%%%

%%%%%%%%%%%%%%%%%%%%%%%%%%%%%%%%%%%%%%%%%%%%%%%%%%%%%%%%%%%%%%%%%%%%%%
%\begin{multicols}{2} %\narrowtext
%%%%%%%%%%%%%%%%%%%%%%%%%%%%%%%%%%%%%%%%%%%%%%%%%%%%%%%%%%%%%%%%%%%%%%

%%%%%%%%%%%%%%%%%%%%%%%%%%%%%%%%%%%%%%%%%%%%%%%%%%%%%%%%%%%%%%%%%%%%%%
%\begin{multicols}{2} %\narrowtext
%%%%%%%%%%%%%%%%%%%%%%%%%%%%%%%%%%%%%%%%%%%%%%%%%%%%%%%%%%%%%%%%%%%%%%

%\end{multicols} %\widetext
%%%%%%%%%%%%%%%%%%%%%%%%%%%%%%%%%%%%%%%%%%%%%%%%%%%%%%%%%%%%%%%%%%%%%%%%%%
%%%%%%%%%%%%%%%%%%%%%%%%%%%%%%%%%%%%%%%%%%%%%%%%%%%%%%%%%%%%%%%%%%%%%%%%%%

%%%%%%%%%%%%%%%%%%%%%%%%%%%%%%%%%%%%%%%%%%%%%%%%%%%%%%%%%%%%%%%%%%%%%%
%%%%%%%%%%%%%%%%%%%%%%%%%%%%%%%%%%%%%%%%%%%%%%%%%%%%%%%%%%%%%%%%%%%%

%%%%%%%%%%%%%%%%%%%%%%%%%%%%%%%%%%%%%%%%%%%%%%%%%%%%%%%%%%%%%%%%%%%
%\begin{multicols}{2}%\narrowtext
%%%%%%%%%%%%%%%%%%%%%%%%%%%%%%%%%%%%%%%%%%%%%%%%%%%%%%%%%%%%%%%%%%%%%%

%%%%%%%%%%%%%%%%%%%%%%%%%%%%%%%%%%%%%%%%%%%%%%%%%%%%%%%%%%%%%%%%%%%%%%

%%%%%%%%%%%%%%%%%%%%%%%%%%%%%%%%%%%%%%%%%%%%%%%%%%%%%%%%%%%%%%%%%%%%%%

%\end{multicols} %\widetext

%\end{multicols} %\widetext
%%%%%%%%%%%%%%%%%%%%%%%%%%%%%%%%%%%%%%%%%%%%%%%%%%%%%%%%%%%%%%%%%%%%%%

%\begin{multicols}{2}%\narrowtext
%%%%%%%%%%%%%%%%%%%%%%%%%%%%%%%%%%%%%%%%%%%%%%%%%%%%%%%%%%%%%%%%%%%%%%

%%%%%%%%%%%%%%%%%%%%%%%%%%%%%%%%%%%%%%%%%%%%%%%%%%%%%%%%%%%%%%%%%%%%%%
%%%%%%%%%%%%%%%%%%%%%%%%%%%%%%%%%%%%%%%%%%%%%%%%%%%%%%%%%%%%%%%%%%%%

%%%%%%%%%%%%%%%%%%%%%%%%%%%%%%%%%%%%%%%%%%%%%%%%%%%%%%%%%%%%%%%%%%%%%%
%\begin{multicols}{2}%\narrowtext
%%%%%%%%%%%%%%%%%%%%%%%%%%%%%%%%%%%%%%%%%%%%%%%%%%%%%%%%%%%%%%%%%%%%%%

%%%%%%%%%%%%%%%%%%%%%%%%%%%%%%%%%%%%%%%%%%%%%%%%%%%%%%%%%%%%%%%%%%%%%%
%\begin{multicols}{2}%\narrowtext
%%%%%%%%%%%%%%%%%%%%%%%%%%%%%%%%%%%%%%%%%%%%%%%%%%%%%%%%%%%%%%%%%%%%%%

%%%%%%%%%%%%%%%%%%%%%%%%%%%%%%%%%%%%%%%%%%%%%%%%%%%%%%%%%%%%%%%%%%%%%%
%%%%%%%%%%%%%%%%%%%%%%%%%%%%%%%%%%%%%%%%%%%%%%%%%%%%%%%%%%%%%%%%%%%%
%%%%%%%%%%%%%%%%%%%%%%%%%%%%%%%%%%%%%%%%%%%%%%%%%%%%%%%%%%%%%%%%%%%%%%
%%%%%%%%%%%%%%%%%%%%%%%%%%%%%%%%%%%%%%%%%%%%%%%%%%%%%%%%%%%%%%%%%%%%

\end{document}